\documentclass[lettersize,journal]{IEEEtran}
\usepackage{generic}
\usepackage{multirow}
\usepackage{amsmath,amssymb,amsfonts}
\usepackage{algorithmic}
\usepackage{graphicx}
\usepackage{algorithm,algorithmic}
\usepackage{threeparttable}
\usepackage{makecell}
\usepackage{subcaption}
\usepackage{enumitem}
\usepackage{xspace}
\usepackage{xcolor}
\usepackage{textcomp}
\usepackage{hyperref}
\usepackage{amsthm}
\newcommand{\schemeName}{{MedExChain}\xspace}
\usepackage{multicol}
\usepackage{caption}
\newtheorem{definition}{Definition}

\usepackage[backend=biber, style=ieee, sorting=none]{biblatex}
\begin{document}

\title{MedExChain: Enabling Secure and Efficient PHR Sharing Across Heterogeneous Blockchains}

\author{{Yongyang Lv, Xiaohong Li, Kui Chen, Zhe Hou, Guangdong Bai, Ruitao Feng$^{\ast}$}
\thanks{This work was supported by National Key Research and Development Program of China (2021YFF1201102). }
\thanks{Y. Lv, X. Li, K. Chen are with the College of Intelligence and Computing, Tianjin University, China. 
(e-mail: \{lvyongyang, xiaohongli, chen\_kui\}@tju.edu.cn)}
\thanks{Z. Hou is with the School of Information and Communication Technology, Griffith University, Australia. (e-mail: z.hou@griffith.edu.au)}
\thanks{G. Bai is with the School of Electrical Engineering and Computer Science, The University of Queensland, Australia. (e-mail: g.bai@uq.edu.au)}
\thanks{R. Feng is with the Faculty of Science and Engineering, Southern Cross University, Australia.
(e-mail: ruitao.feng@scu.edu.au)}
\thanks{*Corresponding author}
\thanks{Copyright (c) 20xx IEEE. Personal use of this material is permitted. However, permission to use this material for any other purposes must be obtained from the IEEE by sending a request to pubs-permissions@ieee.org.}}



\AtBeginEnvironment{abstract}{\bfseries}
\AtBeginEnvironment{IEEEkeywords}{\bfseries}
\renewcommand{\thesection}{
    \Roman{section}
}

\maketitle
\begin{abstract}
With the proliferation of intelligent healthcare systems, patients' Personal Health Records (PHR) generated by the Internet of Medical Things (IoMT) in real-time play a vital role in disease diagnosis. The integration of emerging blockchain technologies significantly enhanced the data security inside intelligent medical systems. However, data sharing across different systems based on varied blockchain architectures is still constrained by the unsolved performance and security challenges. This paper constructs a cross-chain data sharing scheme, termed MedExChain, which aims to securely share PHR across heterogeneous blockchain systems. The MedExChain scheme ensures that PHR can be shared across chains even under the performance limitations of IoMT devices. Additionally, the scheme incorporates Cryptographic Reverse Firewall (CRF) and a blockchain audit mechanism to defend against both internal and external security threats. The robustness of our scheme is validated through BAN logic, Scyther tool, Chosen Plaintext Attack (CPA) and Algorithm Substitution Attack (ASA) security analysis verification. Extensive evaluations demonstrate that MedExChain significantly minimizes computation and communication overhead, making it suitable for IoMT devices and fostering the efficient circulation of PHR across diverse blockchain systems. 
\end{abstract}


\section{Introduction}

With the rapid development of the Internet of Things (IoT), its advantages in remote control, real-time monitoring, and data collection have been widely recognized and applied in intelligent healthcare systems \cite{32}. These systems can utilize various Internet of Medical Things (IoMT) devices to collect patients' Personal Health Records (PHR) in real-time, providing strong support for health monitoring, early disease diagnosis, and decision-making assistance \cite{37}, \cite{44}, such as CrescereMed \cite{[67]} and Hashed Health \cite{[68]}. PHRs contain sensitive physiological data and medical history, making them highly private and confidential \cite{22}. However, due to the current insecure sharing mechanisms and unclear data ownership, intelligent healthcare systems that collect numerous PHRs have become isolated data islands, preventing the sharing of PHRs across different systems and limiting the full utilization of their value \cite{1}, \cite{2}, \cite{36}.\par
Blockchain technology \cite{31}, with its decentralized and tamper-resistant attributes, is often employed as a trusted entity within intelligent healthcare systems to safeguard user data and privacy \cite{36}, \cite{30}, such as Fortified-Chain \cite{33}, MEdge-Chain \cite{34}, and BCHealth \cite{35}. Given the storage constraints of IoMT devices, numerous studies opt to store data in the InterPlanetary File System (IPFS) while maintaining data indexes on the blockchain \cite{1}, \cite{10}, \cite{41}, \cite{46}. However, existing research predominantly focuses on data sharing within systems operating on the same blockchain \cite{32}, \cite{1}, \cite{2}, \cite{36}, \cite{10}, \cite{3}, \cite{6}, \cite{7}, \cite{8}, \cite{9}, with limited consideration for scenarios involving systems distributed across multiple blockchains \cite{73}. Cross-chain data sharing, which operates across diverse cryptographic systems, is inherently susceptible to a range of security threats \cite{69}. Among these, the Algorithm Substitution Attack (ASA) represents a covert backdoor strategy that exploits Trojan horse mechanisms to compromise the integrity of algorithmic processes \cite{71}, \cite{72}. This vulnerability can result in the unauthorized disclosure of sensitive shared data, posing significant risks, particularly in medical contexts where such breaches can have severe consequences.\par
Currently, three main encryption algorithms can ensure the security of data sharing \cite{75}. Attribute-Based Encryption (ABE) algorithm \cite{32} allows for one-to-many data sharing and improves scheme efficiency, but it relies on an attribute center to allocate attributes, which is unsuitable for cross-chain data sharing scenarios \cite{8}, \cite{4}, \cite{23}. Searchable Encryption (SE) algorithm \cite{36} enables safe and efficient retrieval of ciphertext in the IPFS based on its index, but it can only encrypt the data index, not the data itself \cite{9}, \cite{21}. The Proxy Re-Encryption (PRE) algorithm \cite{41} is widely used in existing data sharing schemes \cite{11}, \cite{5}, \cite{13}, \cite{14}. Its strength lies in enabling a semi-trusted agent to transform the ciphertext created by the data owner into ciphertext that the data user can decrypt using their own private key, bypassing the need for complex steps such as downloading, decrypting, and re-encrypting data, thus simplifying the data sharing process. However, the cryptographic systems among different medical institutions can vary significantly. The encryption algorithms in \cite{8}, \cite{9}, \cite{4}, \cite{11}, \cite{5}, \cite{13}, \cite{14} assume uniform cryptographic mechanisms, making them unsuitable for real medical scenarios. Additionally, these schemes are not designed for IoMT devices with low storage and computational capabilities \cite{4}, \cite{23}, \cite{21}, \cite{11}, \cite{13}, \cite{14}.\par
\begin{figure}[htbp]
\centering
\includegraphics[width=0.48\textwidth]{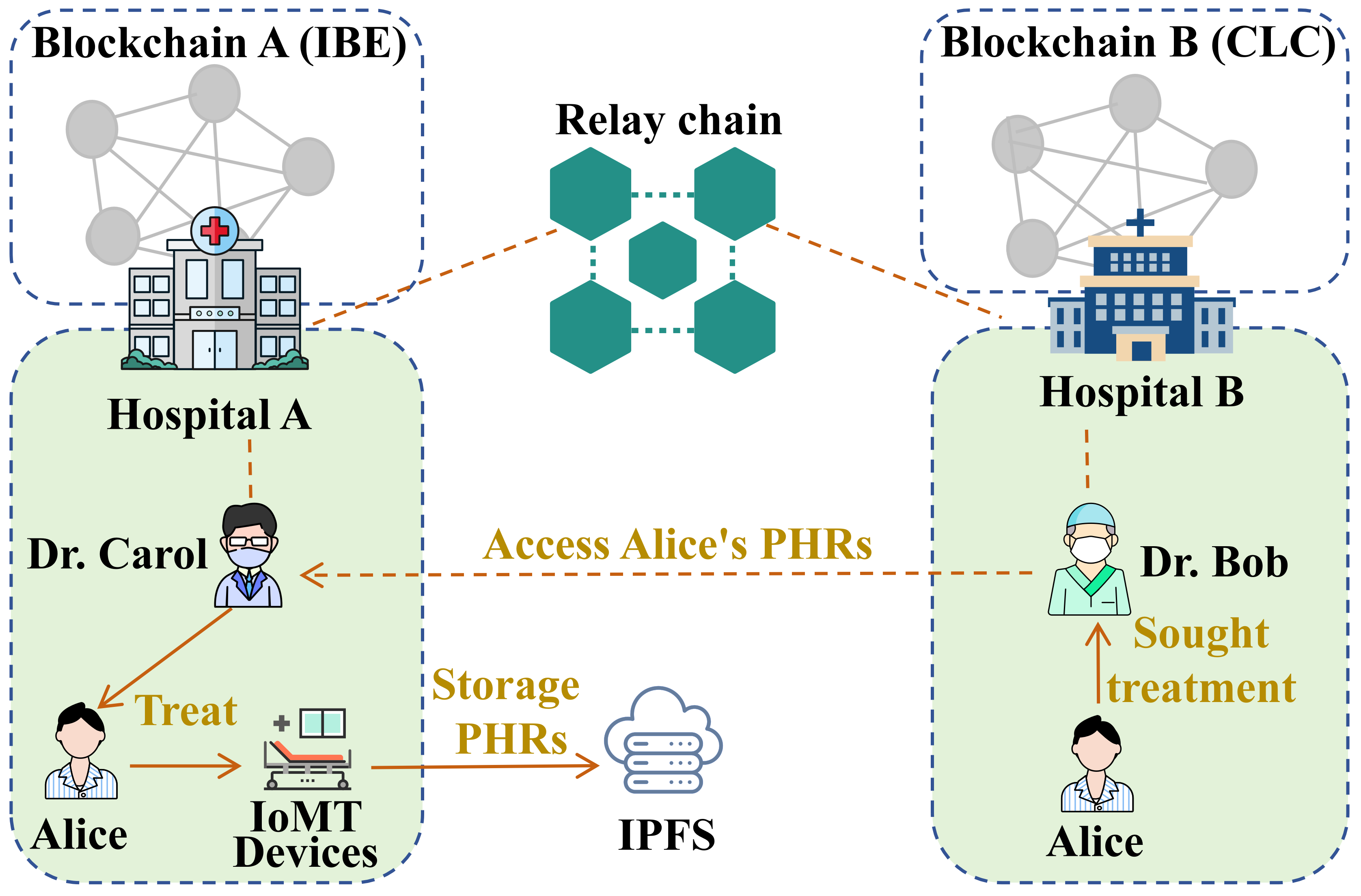}
\caption{The model of sharing PHR across chains}
\label{pic1.1}
\end{figure}
We demonstrated the process of sharing PHR according to two real-world medical institutions based on blockchain, as shown in Figure \ref{pic1.1}. Blockchain A is designed to utilize the Identity-Based Encryption (IBE) cryptosystem, whereas Blockchain B employs the Certificate-Less Cryptography (CLC) cryptosystem. Consider Alice, who was hospitalized at Hospital A, which operates on Blockchain A. During her stay, her PHR was collected in real-time via the IoMT devices of Hospital A. Due to the limited storage and computational capacity of these devices, Alice's PHR was encrypted using her private key and subsequently stored on the IPFS of Hospital A \cite{1}, \cite{10}, \cite{41}, \cite{46}. Later, when Alice sought treatment at Hospital B, which operates on Blockchain B, Dr. Bob required access to the PHR that Alice had previously generated at Hospital A. Since Hospitals A and B are on different blockchain systems, this scenario requires cross-chain PHR sharing. The main challenges in sharing PHRs across chains are:
\begin{enumerate}[leftmargin=*] 
\item [(1)] \textbf{Blockchain Systems are Heterogeneous:} Different blockchains typically employ distinct encryption mechanisms, making it challenging for data users to decrypt information that was encrypted by data owners on a different blockchain \cite{37}, \cite{24}.
\item [(2)] \textbf{The Computational Overhead Increases:} The necessity of converting ciphertext between distinct cryptographic mechanisms introduces computational overhead, leading to communication delays and adversely impacting the overall system efficiency \cite{69}.
\item [(3)] \textbf{Limited Performance of IoMT Devices:} Given the constrained storage and computational capabilities of IoMT devices, performing complex calculations is challenging. PHRs are frequently encrypted and stored in the IPFS, necessitating intricate operations for sharing \cite{10}, \cite{7}, \cite{12}.
\item [(4)] \textbf{Security Threats:} The sharing of PHR is susceptible not only to external threats but also to internal attacks, such as ASA \cite{11}, \cite{12}, \cite{40}, \cite{70}.
\end{enumerate}


To address these challenges, our previous work \cite{10822679} introduced a cross-chain data sharing mechanism leveraging the proxy re-encryption algorithm, enabling the sharing of PHR ciphertexts stored in IPFS between IBE and CLC systems. In contrast, the scenario addressed in this paper is more intricate and demanding. Specifically, we consider an IoMT setting where two medical institutions operating on heterogeneous blockchains require the sharing of PHR data from IoMT devices. While our previous work resolved the issue of differing encryption mechanisms within blockchain systems, this manuscript focuses on overcoming the constraints of low computational and storage capacities of IoMT devices and resisting backdoor attacks during data sharing. The specific improvements include:
\begin{enumerate}[leftmargin=*] 
\item [(1)]\textbf{ Enhanced Security:} To address the issue of ASA in heterogeneous blockchains, we integrate the Cryptographic Reverse Firewall (CRF) \cite{12}, \cite{71}, \cite{72}, which fortifies resistance to backdoor attacks and augments the overall security of the scheme (Sec \ref{C}).

\item [(2)] \textbf{IoMT Terminal PHR Sharing:} This paper facilitates the sharing of PHRs among different blockchain systems. Real-time PHRs generated by IoMT devices are encrypted and stored in IPFS. When data sharing is necessary, IoMT devices with constrained storage and computational capabilities can still facilitate PHR sharing through smart contracts (Overview in Sec\ref{SO}, with details in Sec\ref{C}).

\item [(3)] \textbf{Comprehensive Security Proof:} Whereas our previous work employed Chosen-Plaintext Attack (CPA) security analysis to validate the scheme's security, this paper added the security proof under ASA model, and combined two other proof methods: BAN logic and the \emph{Scyther} tool. These methods are introduced in Sec\ref{RW}, with security goals (Sec\ref{sec:3.2}), security assumptions (Sec\ref{sec:3.3}) and security model (Sec\ref{sec:5.5.4}) defined. Detailed proofs in Sec\ref{sec:5.3},\ref{sec:5.5.5} and\ref{sec:5.2} demonstrate that the MedExChain scheme ensures data confidentiality and integrity, effectively resisting both internal and external attacks.
\item [(4)] \textbf{More Comprehensive Experiments:} Our previous experiments compared computational and communication overheads across three references. This paper expands the comparison to five references, introduces a new blockchain testing dimension, and refines the experimental content concerning computational and communication overheads (Sec \ref{PA}).
\end{enumerate}

In summary, we introduce MedExChain\xspace, a cross-chain sharing scheme for PHR. In this scheme, PHRs generated in real-time are encrypted and stored on the IPFS. When data sharing is required, IoMT devices with constrained storage and computational resources can facilitate PHR sharing by utilizing smart contracts. Additionally, we integrate a CRF and a blockchain audit mechanism into the scheme, ensuring protection against ackdoor attack and preventing information leakage. To evaluate the feasibility of the MedExChain\xspace scheme for cross-chain PHR sharing, we first validate the correctness of the proxy re-encryption algorithm by confirming whether the data user can successfully decrypt the re-encrypted ciphertext. Secondly, we substantiate the high security of our scheme through three distinct security proof methodologies. Finally, we evaluate and contrast the computational and communication overheads of our scheme against other comparable schemes. Additionally, we implement these schemes within a cross-chain system to measure system throughput and latency. We make the following main contributions.


\begin{enumerate}[leftmargin=*] 
    \item [(1)] The MedExChain scheme is proposed, through the improved PRE algorithm, enables secure sharing of PHRs among IoMT devices operating in heterogeneous blockchain systems. This approach effectively addresses the constraints imposed by the limited storage and computational capabilities of these devices.
    \item [(2)] This scheme incorporates a CRF to enhance protection against backdoor attack. Through rigorous analysis using BAN logic, CPA, ASA security analysis, and the \emph{Scyther} tool, we demonstrate that the scheme ensures data confidentiality and integrity, and is resilient to both internal and external threats.
    \item [(3)] Our experimental evaluation, conducted from three analytical perspectives, reveals that the MedExChain\xspace scheme outperforms five comparable references. Notably, it exhibits superior performance in the $ReKeyGen$ and $ReEnc$ stages, achieving the lowest communication overhead per data unit.
\end{enumerate}


\section{Related Works}\label{RW}

{\subsection{Medical Data Sharing Based on ABE}}
 In the context of medical data sharing, the ABE algorithm enables fine-grained access control based on attributes, facilitating one-to-many data sharing capabilities. Quan et al. \cite{2} proposed a reliable medical data-sharing framework in an edge computing environment, addressing the challenges of real-time, multi-attribute authorization in ABE through a blockchain-based distributed attribute authorization strategy (DAA). Hong et al. \cite{46} developed a system that integrates ABE with blockchain to manage Electronic Health Records (EHRs) with fine-grained access control tailored to patients. To mitigate storage costs, the system employs a chameleon hash function to determine the storage addresses of IPFS files. Wang et al. \cite{8} proposed a decentralized electronic medical record-sharing framework called MedShare, which designed a constant-size ABE scheme to achieve fine-grained access control. Zhao et al. \cite{23} proposed a large-scale, verifiable and privacy-preserving dynamic fine-grained access control scheme based on attribute-based proxy re-encryption. While these medical data sharing schemes offer good security and performance, they do not address the issue of cross-chain medical data sharing. Xu et al. \cite{42} introduced a novel privacy-preserving medical data sharing scheme that leverages blockchain and ABE to implement an authorization mechanism. This approach transcends system boundaries, enabling data sharing across multiple medical institutions.
{\subsection{Medical Data Sharing Based on SE}}
In the context of medical data sharing, the SE algorithm facilitates key search capabilities, thereby enabling precise data sharing. Chen et al. \cite{36} introduced BPVSE, a novel verifiable dynamic cloud-assisted EHR scheme that enables users to publicly verify search results returned by the cloud without the need for a trusted authority. BPVSE employs a novel hidden data structure to support dynamic datasets while ensuring forward and backward security.  Liu et al. \cite{7} combined ABE and SE to propose a multi-keyword search-based data-sharing scheme, providing comprehensive privacy protection and efficient ciphertext retrieval for electronic medical records. Banik et al. \cite{9} utilized public key encryption with keyword search (PEKS) technology to design a federated blockchain with preselected users, achieving data security, access control, privacy protection, and secure search. Jiang et al. \cite{21} proposed a cross-domain encrypted exchange service that seamlessly integrates traditional public key encryption with identity-based encryption. This approach allows for secure data search (outsourcing) post-encryption, ensuring data integrity and maintaining query confidentiality.

{subsection{Medical Data Sharing Based on PRE}}
In the context of medical data sharing, blockchain is commonly considered a reliable entity for facilitating the exchange of medical data across different systems, thereby enabling cross-system data sharing. Liu et al. \cite{43} alleviated the substantial data storage burden of medical blockchain by employing an "on-chain and off-chain" approach. Pei et al. \cite{45} introduced a secure data sharing scheme called PRE-IoMT. In this scheme, an identity hash is incorporated during the key generation stage to bind public keys with user identities, thereby enhancing the security of data sharing within PRE-IoMT. Sur et al. \cite{16} introduced the concept of certificateless PRE, providing a precise definition of secure certificateless PRE schemes. Ge et al. \cite{15} proposed a verifiable and fair attribute-based PRE scheme (VF-ABEPRE), using message-locked encryption technology to ensure that the same plaintext corresponds to the same re-encrypted ciphertext. Zhou et al. \cite{12} designed an identity-based PRE scheme with a cryptographic reverse firewall (IBPRE-CRF), offering security against chosen-plaintext attacks and resistance to exfiltration attacks. Mizuno et al. \cite{17} and Deng et al. \cite{18} proposed using PRE to convert ABE ciphertexts to IBE ciphertexts. While existing research has extensively explored ciphertext conversion within the same cryptographic system, there is still room for improvement in the security and performance of PRE schemes across different cryptographic systems.
\section{Preliminaries}\label{P}
{This section introduces the concepts of bilinear pairings and Cryptographic Reverse Firewall (CRF), and also introduces Ban logic and \emph{Scyther} tool.} 
\subsection{Bilinear Pairing}
Let $G_1$ and $G_2$ be two multiplication groups of order prime $q$, with $g$ as the generator of $G_1$. A bilinear pairing $e:G_1 \times G_1\rightarrow G_2$ satisfies the following properties:
\begin{enumerate}[leftmargin=*] 
\item [(1)] Bilinearity: For $\forall\left(g_1,g_2\right)\in G_1$, $\forall\left(a,b\right)\in Z_q^\ast$, it must hold that $e\left(g_1^a,g_2^b\right)=e\left(g_1,g_2\right)^{ab}$.
\item [(2)] Non-degeneracy: For $\exists\left(g_1,g_2\right)\in G_1$ and $1_{G_2}$ be the identity element of $G_2$, there have $e\left(g_1,g_2\right)zhizai1_{G_2}$.
\item [(3)] Computability: For $\forall\left(g_1,g_2\right)\in G_1$, there exists an effective algorithm to compute $e\left(g_1,g_2\right)$.
\end{enumerate}

\begin{figure*}[!t]
    \centering
    \includegraphics[width=17.6cm]{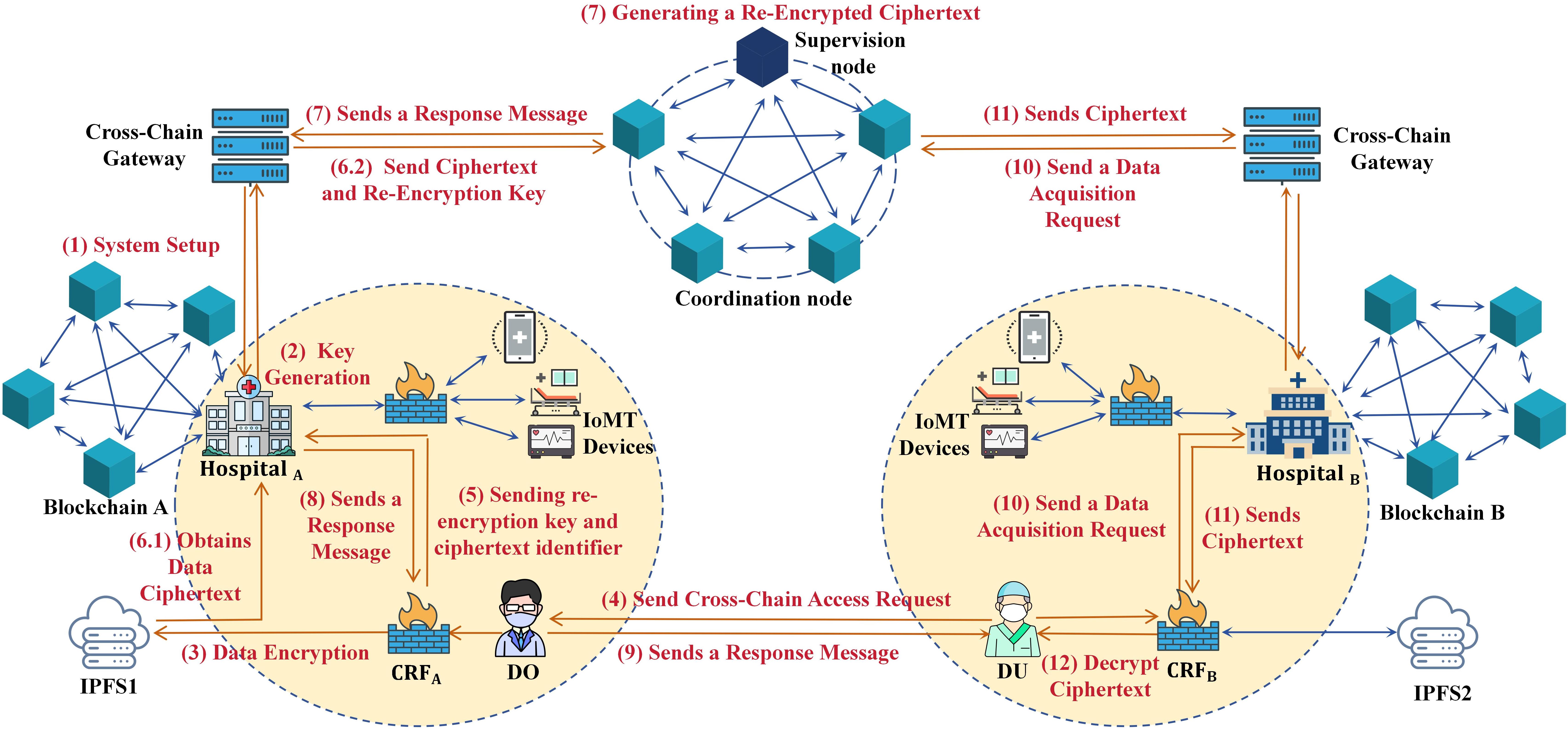}
    \caption{The \schemeName scheme model}
    \label{structure}
\end{figure*}

{\subsection{Cryptographic Reverse Firewalls}
In \cite{19}, let $W$ be a CRF, $P=\left(receive,next,output\right)$ be a party, we can say $W$ is a CRF for $P$ if it meets the following properties. Here, $\sigma$ is an initial public parameter, $m$ is the transmitted message. Define $W\circ P$ as follows:
\vspace{-1ex}
\begin{gather*}
    W\circ P {:=}(receive_{W\circ P}\left(\sigma,m\right)=receive_P\left(\sigma,W\left(m\right)\right)\\
    next_{W\circ P}\left(\sigma\right) =W\left(next_P\left(\sigma\right)\right)\\
    output_{W\circ P}\left(\sigma\right) =output_P\left(\sigma\right))
\end{gather*}
 A qualified CRF needs to satisfy the following properties:
\begin{enumerate}[leftmargin=*] 
\item [(1)] Functionality-maintaining. If the user’s computer operates correctly, the CRF will not compromise the functionality of the cryptographic algorithms.
\item [(2)] Weak security-preserving. Regardless of how the user’s computer is affected by an attacker, the use of the CRF will remain as secure as the correct execution of cryptographic algorithms.
\item [(3)] Exfiltration-resistant. No matter how to run the user’s computer, the CRF will prevent the computer from leaking confidential information.
\end{enumerate}

\subsection{BAN Logic}
BAN logic is a modal logic grounded in belief employed to formally assess the security of protocols. In Sec \ref{sec:5.3}, this paper employs BAN logic to assess the security of the MedExChain scheme. The fundamental principle of BAN logic involves deducing the final beliefs of the protocol participants from their initial beliefs through a series of logical inference rules. To gain a comprehensive understanding of these inference rules, please consult reference \cite{25}. To evaluate a protocol using BAN logic, one must initially idealize the protocol, transforming its messages into BAN logic formulae. Subsequently, based on the specific context, appropriate assumptions are made, and through the application of logical reasoning rules, it is determined whether the protocol fulfills its intended objectives.\par

\subsection{Scyther Tool}
 \emph{Scyther} tool \cite{26} is a formal verification tool utilized for analyzing and validating the security attributes of protocols, encompassing confidentiality, authenticity, and integrity. It incorporates the Security Protocol Description Language (SPDL) to facilitate the verification of protocol design specifications and anticipated security assumptions. \emph{Scyther} efficiently scans and detects potential security vulnerabilities within the protocol based on predefined assumptions.
}
 
{\section{System Overview}\label{SO}
This section introduces the system model, workflow, security goals and threat model of the MedExChain\xspace scheme. }

\subsection{System Model}
In the \schemeName scheme, we assume that there is a node, ${\text{Hospital}_\text{A}}$, in Blockchain A, which uses IBE as its cryptographic system. Similarly, there is a node, ${\text{Hospital}_\text{B}}$, in Blockchain B, which uses CLC as its cryptographic system. Blockchains A and B are connected via ${\text{Relay}}$ to cross-chain gateways. This scheme assumes that a data user in ${\text{Hospital}_\text{B}}$ needs to access some PHRs from a data owner in ${\text{Hospital}_\text{A}}$. The \schemeName scheme model is shown in Figure \ref{structure}, and the scheme includes the following entities:

\subsubsection{Hospital (${\text{Hospital}_\text{i}}$)} A node in the blockchain that generates keys for users within the chain. It is a trusted entity with the highest authority level in the node and manages transactions within the node.

\subsubsection{Data Owner (DO)} A user in ${\text{Hospital}_\text{A}}$ who owns the PHR. The DO can be any type of IoMT device with limited computing and storage resources.

\subsubsection{Data User (DU)} A user in ${\text{Hospital}_\text{B}}$ who can be a patient, doctor, researcher, or any other person needing to use PHR.

\subsubsection{Relay-Chain (Relay)} Consists of nodes with different functions. The computing node provides computing power and is responsible for the calculation of re-encrypted ciphertext. The supervision node and the coordination node use blockchain smart contracts to authenticate and register the blockchain connected to the ${\text{Relay}}$, and to confirm and audit intra-chain transactions.

\subsubsection{Interplanetary File System (${\text{IPFS}_\text{i}}$)} A semi-trusted distributed database responsible for storing PHRs to reduce the storage burden on IoMT devices.

\subsubsection{Cryptographic Reverse Firewal (${\text{CRF}_\text{i}}$)} Deployed between user terminal equipment and blockchain nodes. If an attacker intercepts a message through a backdoor, the CRF can prevent the attacker from knowing the exact content of the user's message.

{\subsection{System Workflow}

In the \textbf{System Setup} stage, (1) Blockchains A and B must authenticate and register with the ${\text{Relay}}$ through a cross-chain gateway, and generate system parameters. \par
In the \textbf{Key Generation} stage, (2) Blockchains A and B generate keys for users in their respective chains. \par
In the \textbf{Data Encryption} stage, (3) during routine operations, the DO encrypts relevant medical data with its private key, stores the encrypted data on the IPFS, and saves the ciphertext identifier $Data_1$ locally. \par
In the \textbf{Data Request} stage, (4) when the DU requires access to the DO's medical data, it sends a cross-chain access request message to the DO. (5) Upon receiving the request, the DO verifies it, and upon successful verification, the DO computes a re-encryption key $RK$, sending the $Data_1$ along with the $RK$ to ${\text{Hospital}_\text{A}}$. (6.1) ${\text{Hospital}_\text{A}}$ obtains data ciphertext $M_1$ in IPFS according to the $Data_1$, (6.2) then sends the $M_1$ along with the $RK$ to the ${\text{Relay}}$. (7) The ${\text{Relay}}$ re-encrypts the $M_1$ using the $RK$, generating a re-encrypted ciphertext $M_2$ that DU can decrypt, along with an re-encrypted ciphertext identifier $Data_2$, and sends $Data_2$ to ${\text{Hospital}_\text{A}}$. (8) ${\text{Hospital}_\text{A}}$ then sends a response message containing the $Data_2$ to the DO. (9) The DO forwards a response message with $Data_2$ to the DU. \par
In the \textbf{Data Acquisition} stage, (10) the DU requests the $M_2$ from the ${\text{Relay}}$. (11) The ${\text{Relay}}$ retrieves the $M_2$ based on the $Data_2$ and sends it to the DU. (12) The DU can decrypt the $M_2$ using its private key to obtain the relevant medical data.
\subsection{Baseline}
\subsubsection{Password Assumption} The encryption algorithms in schemes \cite{46}, \cite{8},  \cite{42}, \cite{45} typically depend on a uniform encryption mechanism, which is impractical for real-world medical scenarios. The encryption systems of various medical institutions can differ significantly, necessitating encryption algorithms adaptable to diverse systems in the proposed scheme.

\subsubsection{Computational Overhead} The algorithms in schemes \cite{36}, \cite{7}, \cite{15}, \cite{47} are unsuitable for IoMT devices with constrained storage and computing capabilities. The proposed scheme accommodates these limitations by minimizing computational overhead, allowing devices to perform basic computations and manage data encryption or decryption within feasible limits.

\subsubsection{Security Threats} Schemes \cite{46}, \cite{8}, \cite{40} fail to address internal attacks that commonly arise during data sharing. The proposed scheme must mitigate both internal and external threats prevalent in data sharing processes, thereby ensuring data confidentiality and integrity.

\subsection{Security Goals}
\label{sec:3.2}
  \subsubsection{Correctness} Users can use their private keys to decrypt ciphertext correctly. Ensuring the accurate execution of protocols in alignment with established rules and standards. (The proof in Sec.\ref{sec:5.1} and \ref{sec:5.3})
    \subsubsection{Confidentiality} Users' keys, data and other information should be protected from enemy attacks. (The proof in Sec.\ref{sec:5.5.5} and \ref{sec:5.2})
\subsubsection{Integrity} It can prove that the message content has not been modified during transmission. (The proof in Sec.\ref{sec:5.5.5} and \ref{sec:5.2})

\subsection{Security Assumptions and Threat Model}
\label{sec:3.3}
\subsubsection{Assumptions Regarding Cryptographic Algorithms}
We assume that the cryptographic algorithm employed is secure, implying that without knowledge of the correct key, an adversary is incapable of decrypting the message.

\subsubsection{Assumptions Regarding Entities}
We consider DO, ${\text{Hospital}_\text{i}}$, and ${\text{CRF}_\text{i}}$ to be entirely reliable, ensuring that attackers cannot eavesdrop, intercept, or manipulate the communication channels between these entities. ${\text{IPFS}_\text{i}}$ and Relay are classified as semi-trusted, adhering to protocol requirements but exhibiting curiosity about message content. DU is deemed untrustworthy, potentially under the control of a malicious entity.
\subsubsection{Assumptions Regarding Adversaries}
An outside adversary may attempt to extract sensitive information from the sender's encrypted ciphertext within the ${\text{IPFS}_\text{i}}$, Relay, or communication channel.
}

\begin{figure*}[htbp]
 \centering
 \includegraphics[width=1\textwidth]{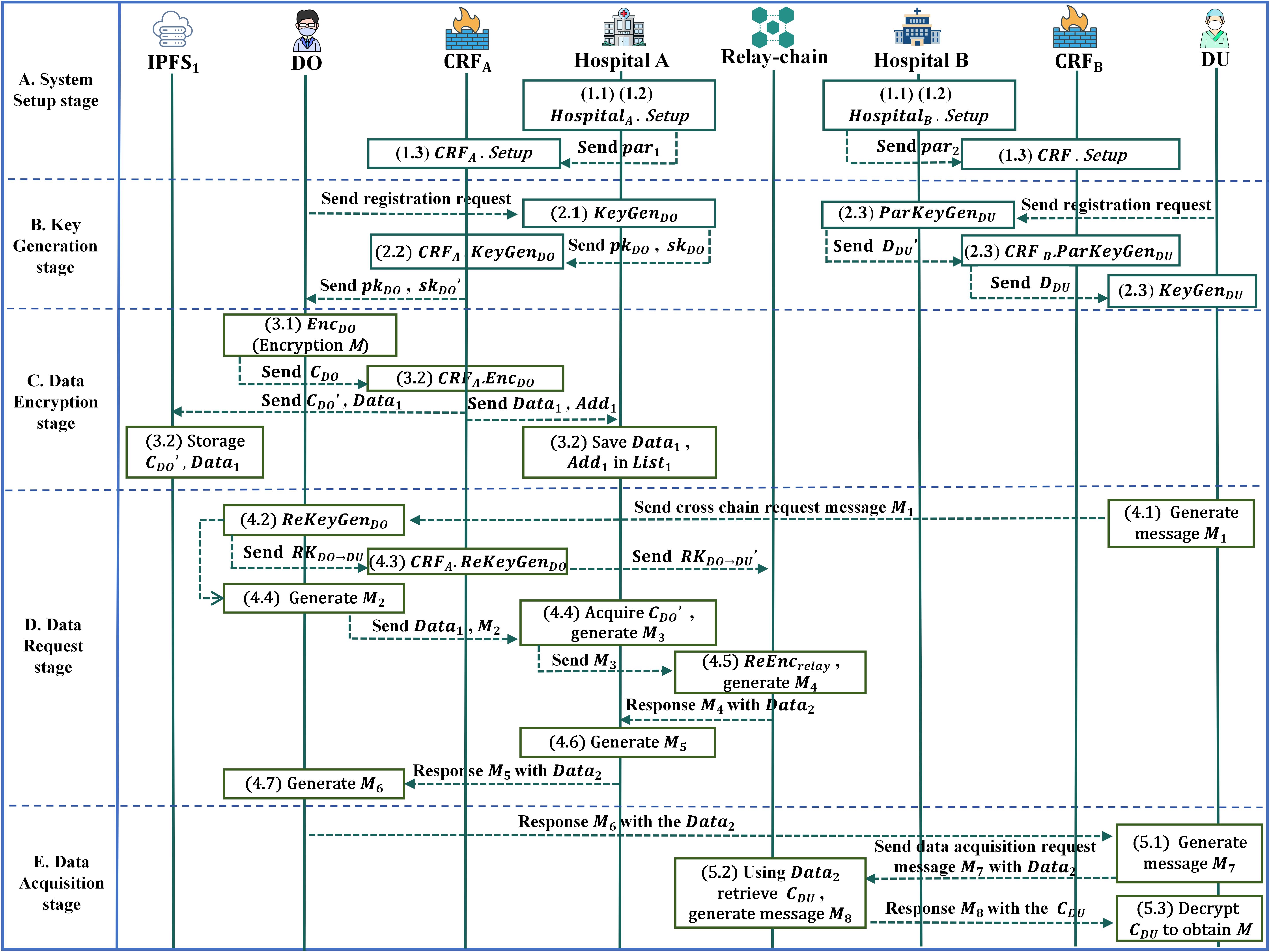}
\caption{Workflow of MedExChain scheme}
 \label{liuchengtu}
 \end{figure*}

{\section{The \schemeName Scheme Construction}\label{C}
This paper improves the proxy re-encryption algorithm in \cite{12} and constructs the \schemeName scheme. This section introduces the details of the scheme's implementation. Fig.~\ref{liuchengtu} shows the workflow of the MedExChain scheme, the symbol descriptions in the scheme are shown in Table \ref{tab:G}, and the details of each stage are described as follows:}

\begin{table}
\centering
\caption{{\MakeUppercase {Description of symbols }}}
\label{tab:G}
\begin{tabular}{|p{1.5cm}|p{5.4cm}|}
    
    \hline
    \multirow{4}{*}{}
     \textbf{Symbol}  &  \textbf{Description}\\ 
    \hline
     $par_i$  & System parameters\\
    \hline
     $ID_{i}$  & User $i$'s identity\\
    \hline
     $D_{i}$  & User $i$'s partial private key\\
    \hline
     $sk_i$  & User $i$'s private key\\
    \hline
    $pk_i$ & User $i$'s public key \\
    \hline
    $M$ & Messages containing PHR \\
    \hline
    $C_{DO}$ & Original ciphertext\\
    \hline
    $C_{DO}\prime$ & The ciphertext processed by $CRF_A$\\
    \hline
    $RK_{DO \to DU}$ & Re-encryption key\\
    \hline
    $RK_{DO \to DU\prime}$ &The re-encryption key processed by $CRF_A$\\
    \hline
    $C_{DU}$	& Re-encrypted ciphertext\\
    \hline
    $T_i$ & Timestamp\\ 
    \hline
    $N_i$	& Parameter of keeping session fresh\\
    \hline
\end{tabular}
\end{table}

\subsection{System Setup}
\label{1.1}
At this stage, Blockchains A and B register their systems to generate system parameters. The specific process is as follows:

\begin{enumerate}
\item [(1.1)] Given a security parameter $k$, the ${\text{Hospital}_\text{A}}$ in Blockchain A and the ${\text{Hospital}_\text{B}}$ in Blockchain B select two multiplicative groups $G_1$ and $G_2$ with a prime number $q$, a bilinear pair $e:G_1 \times G_1 \rightarrow G_2$, and two hash functions $H_1:\{0,1\}^*\rightarrow G_1$ and $H_2:G_2\rightarrow G_1$.
\item [(1.2)] ${\text{Hospital}_\text{A}}$ selects $s \in Z_q^*$ as the master private key and calculates the system public key $PK_A=g^s$, where $g$ is the generator of $G_1$. ${\text{Hospital}_\text{A}}$ sends the parameters $par_1=\{G_1,G_2,e,g,PK_A,H_1,H_2\}$ to its cryptographic reverse firewall ${\text{CRF}_\text{A}}$. Similarly, ${\text{Hospital}_\text{B}}$ in Blockchain B randomly selects $y\in Z_q^*$ as the master private key and calculates the system public key $PK_B=g^y$.
{\item [(1.3)] ${\text{CRF}_\text{A}}$ randomly selects $a\in Z_q^*$ as the master key, calculates $PK_A\prime={PK_A}^a=g^{sa}$, and updates the parameters to $par_1\prime=\{G_1, G_2, e, g, PK_A\prime, H_1, H_2\}$. Similarly, ${\text{CRF}_\text{B}}$ randomly selects $b\in Z_q^\ast$ as the master private key, calculates $PK_B\prime={PK_B}^b=g^{yb}$, and updates the parameters to $par_2\prime=\{G_1, G_2, e, g, PK_B\prime, H_1, H_2\}$.}
\end{enumerate}

\subsection{Key Generation}
At this stage, Blockchains A and B generate keys for users in their respective chains. The specific process is as follows:

\begin{enumerate} 
    \label{2.1}
    \item [(2.1)] \textit{\textbf{KeyGen$_{DO}$ ($ID_{DO}, s \to sk_{DO}, pk_{DO}$)}}: Given the DO's identity $ID_{DO}\in\{0,1\}^*$, ${\text{Hospital}_\text{A}}$ generates the user's public key $pk_{DO}=H_1\left(ID_{DO}\right)$ and private key $sk_{DO}=pk_{DO}^s$, and sends $sk_{DO}$ to ${\text{CRF}_\text{A}}$.
    \label{2.2}
    {\item [(2.2)] \textit{\textbf{${\text{CRF}_\text{A}}$-KeyGen$_{DO}$ ($sk_{DO}, a \to sk_{DO}\prime$)}}: Upon receiving ${sk}_{DO}$, ${\text{CRF}_\text{A}}$ uses the master private key $a$ to generate a randomized user's private key $sk_{DO}\prime=sk_{DO}^a=pk_{DO}^{sa}$, and then sends $sk_{DO}\prime$ to the DO.}
    
     \item [(2.3)] \textit{\textbf{KeyGen$_{DU}$ ($ID_{DU}, y, b, r \to sk_{DU}, pk_{DU}$)}}: \\
     a) Given the DU's identity $ID_{DU}\in\{0,1\}^*$, the ${\text{Hospital}_\text{B}}$ generates the user's partial private key $D_{DU}=H_1\left(ID_{DU}\right)^y$. {${\text{Hospital}_\text{B}}$ sends $D_{DU}$ to ${\text{CRF}_\text{B}}$.
    b) Upon receiving $D_{DU}$, ${\text{CRF}_\text{B}}$ uses the master private key $b$ to generate a randomized user's partial private key $D_{DU}\prime=D_{DU}^b=H_1\left(ID_{DU}\right)^{yb}$, and then sends $D_{DU}\prime$ to the DU.
    c) The DU randomly selects $r\in Z_p^\ast$, calculates the private key $sk_{DU}=(D_{DU}{\prime)}^r=H_1\left(ID_{DU}\right)^{ybr}$, and the public key $pk_{DU}=(pk_{DU1},pk_{DU2})=\left(H_1\left(ID_{DU}\right),\ (h_2{\prime)}^r\right)$.}
\end{enumerate}

\subsection{Data Encryption}
At this stage, the DO encrypts PHR and uploads it to ${\text{IPFS}_\text{A}}$ for storage.

\begin{enumerate}
\label{3.1}
     \item [(3.1)] \textit{\textbf{Enc$_{DO}$ ($M, par_1\prime, pk_{DO}, \alpha \to C_{DO}$)}}: The DO selects the message $M$ (containing PHR) to be shared, given $par_1\prime$ and $pk_{DO}$, randomly selects $\alpha\in Z_q^\ast$, and generates the ciphertext  $C_{DO}=(c_1,c_2)$. Then, the DO sends $C_{DO}$ to ${\text{CRF}_\text{A}}$.
 \[
\left\{
\begin{aligned}  
       &c_1=g^\alpha  \\
       &c_2=M\cdot e(PK_A\prime,pk_{DO})^\alpha
\end{aligned}
\right.
\]
\label{3.2}
    \item [(3.2)] \textit{\textbf{${\text{CRF}_\text{A}}$-Enc$_{DO}$ ($C_{DO}, \beta \to C_{DO}\prime$)}}: Upon receiving $C_{DO}$, ${\text{CRF}_\text{A}}$ randomly selects $\beta\in Z_q^\ast$, and generates the randomized ciphertext $C_{DO}\prime=(c_1\prime,c_2\prime,c_3\prime) $. ${\text{CRF}_\text{A}}$ then sends $ C_{DO}\prime$ and its identifier $Data_1$ to ${\text{IPFS}_\text{A}}$ for storage. Simultaneously, ${\text{Hospital}_\text{A}}$ saves the ciphertext identifier $Data_1$ and its address $Add_1$ in the access list $List_1$.
 \[
\left\{
\begin{aligned}  
         &c_1\prime=c_1 \cdot g^\beta \\
         &c_2\prime=c_2 \cdot e(PK_A\prime,pk_{DO})^\beta \\
         &c_3\prime=pk_{DO}^\beta
%
\end{aligned}
\right.
\]
\end{enumerate}
\subsection{Data Request}
At this stage, the DU initiates a cross-chain access request to the DO. Upon successful verification of the request by DO, it instructs ${\text{Hospital}_\text{A}}$ to transmit the re-encryption key and the data ciphertext to the ${\text{Relay}}$. Subsequently, the ${\text{Relay}}$ generates the re-encryption ciphertext. 

\begin{enumerate}
    \item [(4.1)] To access the message $M$ from the DO, the DU must first send a cross-chain access request message $M_1 = \{request_1,pk_{DO},pk_{DU},T_1,N_1\}_{pk_{DO}}$. Here, $request_1$ is the cross-chain access identifier, $T_1$ is the timestamp, and $N_1$ is the nonce to maintain session freshness. The message $M_1$ is forwarded to the DO via the cross-chain gateway.
    \label{4.2}
    \item [(4.2)] \textit{\textbf{ReKeyGen$_{DO}$ ($\lambda, X, sk_{DO}\prime, pk_{DU} \to RK_{DO \to DU}$)}}: Upon receiving the request, the DO verifies the validity of the message and the correctness of the DU's identity. If the verification is successful, the DO randomly selects $\lambda$ and $X$. Then, using its own private key $sk_{DO}\prime$ and the DU's public key $pk_{DU}$, the DO generates the re-encryption key $RK_{DO \to DU} = ( {rk}_1, {rk}_2, {rk}_3 )$ and sends it to ${\text{CRF}_\text{A}}$.
 \[
\left\{
\begin{aligned}   
         &rk_1 = H_2(X) / sk_{DO}\prime \\
         &rk_2 = g^\lambda \\
         &rk_3 = X \cdot e(pk_{DU1}, pk_{DU2})^\lambda
\end{aligned}
\right.
\]
{\label{4.3}
    \item [(4.3)] \textit{\textbf{${\text{CRF}_\text{A}}$-ReKeyGen$_{DO}$ ($RK_{DO \to DU}, \beta \to RK_{DO \to DU}\prime$)}}: Upon receiving $RK_{DO \to DU}$, ${\text{CRF}_\text{A}}$ generates the randomized re-encryption key $RK_{DO \to DU}\prime = (rk_1\prime, rk_2\prime, rk_3\prime)$.
    \[
\left\{
\begin{aligned}
          &rk_1\prime = rk_1 \cdot pk_{DO}^{-\beta} \\
         &rk_2\prime = rk_2 \cdot g^\beta \\ 
          &rk_3\prime = rk_3 \cdot e(pk_{DU1}, pk_{DU2})^\beta
\end{aligned}
\right.
\]}
    \item [(4.4)] The DO then sends the ciphertext identifier $Data_1$ and the cross-chain data sharing permission message $M_2=\{request_2,pk_{DO},pk_{DU},Data_1,RK_{DO \to DU}\prime,T_2,N_2\}$ to ${\text{Hospital}_\text{A}}$. ${\text{Hospital}_\text{A}}$ employs the intelligent contract algorithm \textbf{${\text{Algorithm}_\text{1}}$} to acquire the data ciphertext $C_{DO}\prime$ from ${\text{IPFS}_\text{A}}$ based on $Data_1$. Subsequently, it generates a cross-chain data conversion request message $M_3=\{request_3,pk_{DO},pk_{DU},C_{DO}\prime,RK_{DO \to DU}\prime,T_3,N_3\}$ and transmits it to the ${\text{Relay}}$.
\label{4.5}
    \item [(4.5)] \textit{\textbf{ReEnc$_{relay}$ ($C_{DO}\prime, RK_{DO \to DU}\prime \to C_{DU}$)}}: The ${\text{Relay}}$ generates the re-encrypted ciphertext $C_{DU} = (C_1, C_2, C_3, C_4)$ based on the given $C_{DO}\prime$ and $RK_{DO \to DU}\prime$. Finally, the ${\text{Relay}}$ sends the response message $M_4=\{respond_1,Data_2,T_4,N_4\}$ with the identifier $Data_2$ of $C_{DU}$ to the ${\text{Hospital}_\text{A}}$ through the cross-chain gateway.
    \[
\left\{
\begin{aligned} 
        &C_1 = c_1\prime \\
        &C_2 = c_2\prime \cdot e(C_1, rk_1\prime \cdot c_3\prime)\\
        &C_3 = rk_2\prime \\
       & C_4 = rk_3\prime
\end{aligned}
\right.
\]
    \item [(4.6)] Upon receiving the response message $M_4$, ${\text{Hospital}_\text{A}}$ initially validates the authenticity of the message. Following successful validation, ${\text{Hospital}_\text{A}}$ generates a response message $M_5=\{respond_2,Data_2,T_5,N_5\}$ and transmits it to the DO.
    \item [(4.7)] Upon receiving the message $M_5$, the DO generates a response message $M_6=\{respond_3,Data_2,T_6,N_6\}$ and transmits it to the DU.
    \end{enumerate}

\subsection{Data Acquisition}
At this stage, the DU submits a request to the ${\text{Relay}}$ for data retrieval. Subsequently, DU decrypts the acquired data using its private key.
\begin{enumerate} 
    {\item [(5.1)] Upon receiving the response message $M_6$, the DU first verifies the authenticity of the message. Upon successful verification, it generates a data acquisition request message $M_7=\{request_4,pk_{DO},pk_{DU},Data_2,T_7,N_7\}$ and transmits it to the ${\text{Relay}}$.
    \item [(5.2)] Upon receiving message $M_7$, the ${\text{Relay}}$ retrieves the re-encrypted ciphertext $C_{DU}$ using the identifier $Data_2$, subsequently generating a response message $M_8=\{respond_4,C_{DU},T_8,N_8\}$ which is then transmitted to the DU.}
    \item [(5.3)] \textit{\textbf{Dec$_{DU}$ ($C_{DU}, sk_{DU} \to M$)}}: Upon receiving the response message $M_8$, the DU employs its private key $sk_{DU}$ to compute the ciphertext $X = C_4 / e(C_3, sk_{DU})$, and then calculates $M = C_2 / e(C_1, H_2(X))$ to obtain the message $M$.
\end{enumerate}

\begin{table}[h!]
\centering
\begin{tabular}{p{8cm}}
\hline
\textbf{${\text{Algorithm}_\text{1}}$}\\
\hline
\textbf{Input: }$Data_1$, $M_i$(request, $ID_{DU}$, Timestamp) \\
\textbf{Output: }True/False\\
\textbf{Begin:} \\
\quad 1. ${\text{Hospital}_\text{i}}$ receives message $(Data_1, M_i)$\\
\quad 2. // get the number of DU's access from access list\\
\quad 3. accessCount $\gets$ countOf($List_1$, find($ID_{DU}$))\\
\quad 4. \textbf{if !}accessCount $<$ maxAccessCount \textbf{then}\\
\quad 5. \quad Display ``Access limit reached!''\\
\quad 6. \quad \textbf{return} False\\
\quad 7. // search the address of ciphertexts marked by $Data_1$ in the blockchain\\
\quad 8. $addr_1 \gets$ searchInChain($Data_1$)\\
\quad 9. \textbf{if !}$addr_1$ exists \textbf{then}\\
\quad 10. \quad Display "Target data doesn’t exist"\\
\quad 11. \quad \textbf{return} False\\
\quad 12. // get ciphertext in IPFS based on address $addr_1$\\
\quad 13. $C_{DO}\prime \gets$ searchInIPFS$_\text{1}$($addr_1$)\\
\quad 14. ${\text{Hospital}_\text{1}}$ sends message $(C_{DO}\prime, M_i)$ to RelayChain\\
\quad 15. // record this access\\
\quad 16. $List$.insert($ID_{DU}$, Timestamp, $Data_1$)\\
\quad 17. \textbf{return} True\\
\textbf{end}\\
\hline
\end{tabular}
\end{table}

{\section{Correctness and Safety Analysis}}\label{CSA}
 This section first proves the correctness of the \schemeName scheme. Secondly, we use Ban logic, CPA and ASA security analysis, and \emph{Scyther} tool to prove the security of the \schemeName scheme.

 \subsection{Scheme Correctness Proof}
\label{sec:5.1}
\par
\begin{definition}
 If DO encrypts message $M$ to generate ciphertext $CT$, and the re-encrypted ciphertext is $CT’$, then proxy re-encryption algorithm is correct if $Decrypt_{DU}(CT’) = M$.
\end{definition}
\par
\begin{proof}
 We verify the correctness of the \schemeName scheme by checking if the DU can accurately decrypt the re-encrypted ciphertext $C_{DU} = (C_1, C_2, C_3, C_4)$.
{
\vspace{-1ex}
\begin{align*}
    \frac{C_4}{e\left(C_3,sk_{DU}\right)}&=\frac{rk_3\prime}{e\left(rk_2\prime,sk_{DU}\right)}\\
    &=X \cdot \frac{e(pk_{DU1},pk_{DU2}) ^{\lambda+\beta}}{e(g^{\lambda+\beta},sk_{DU})}\\
    &=X \cdot \frac{e(H_1\left(ID_{DU}\right),g^{ybr}) ^{\lambda+\beta}}{e(g^{\lambda+\beta},H_1\left(ID_{DU}\right)^{ybr})}\\
    &=X
\end{align*}}

It is evident that $X$ can be correctly decrypted by DU.
\vspace{-1ex}
\begin{align*}
    \frac{C_2}{e\left(C_1,H_2\left(X\right)\right)}&= \frac{c_2\prime\cdot e(c_1\prime,rk_1\prime\cdot c_3\prime)}{e(c_1\prime,H_2(X))}\\
    &= M \cdot \frac{e\left(g^{sa},pk_{DO}\right)^{\alpha+\beta}\cdot e(g^{\alpha+\beta},H_2(X))}{e\left(g^{\alpha+\beta},H_2\left(X\right)\right) \cdot e(g^{\alpha+\beta},sk_{DO}\prime)}\\
    &= M \cdot \frac{e\left(g^{sa},pk_{DO}\right)^{\alpha+\beta}}{e\left(g^{\alpha+\beta},sk_{DO}\prime\right)}\\
    &= M
\end{align*} 

Based on the correct decryption of $X$, the DU also correctly decrypts the ciphertext $C_{DU}$ to obtain message $M$.
\end{proof}

\subsection{BAN Logic Proof}{
\label{sec:5.3}
In this section, we conduct a security assessment of the \schemeName scheme's logic using BAN logic. BAN logic \cite{25} is a belief-based modal logic used to establish an idealized protocol model, making reasonable assumptions about specific situations. By applying inference rules to the idealized protocol and assumptions, we can deduce whether the protocol achieves its intended goals.

 \subsubsection{Constructing an Idealized Protocol Model}
 In our idealized protocol model, we consider four entities as principals: DO, DU, ${\text{Hospital}_\text{A}}$ and the ${\text{Relay}}$. Based on the scheme's description, we can construct the following idealized protocol model, divided into several messages:

 Message $M_1$: DU$\to$DO :\\ $\{request_1,pk_{DO},pk_{DU},T_1,N_1,\text{DO}\stackrel{\text{SK}}{\longleftrightarrow}\text{DU}\}_{pk_{DO}} $

 Message $M_2$: $\text{DO}\to$ ${\text{Hospital}_\text{A}}$:\\ $\{request_2,pk_{DO},pk_{DU},RK_{DO \to DU}\prime,T_2,N_2\}$

 Message $M_3$: $ {\text{Hospital}_\text{A}} \to \text{RelayChain}:\\$ $\{request_2,pk_{DO},pk_{DU},RK_{DO \to DU}\prime,T_3,N_2,C_{DO}\prime\}$

 Message $M_4$: $ \text{RelayChain} \to \text{DU}:\\$ $\{respond_1,C_{DU},T_4,N_3\}$

 Message $M_5$: $ \text{DO}\to \text{DU}:\\ \{respond_2,pk_{DO},pk_{DU},T_5,N_4,\text{DO}\stackrel{\text{SK}}{\longleftrightarrow}\text{DU}\}_{pk_{DU}} $

 \subsubsection{Setting Goals}
 Based on the BAN logic language, the protocol's goals are described such that the protocol can resist malicious attacks only when it achieves these predetermined goals. For this protocol, we set the following four goals:

 Goal $G_1$: $\text{DU}|\equiv \text{DO}|\equiv (\text{DU}\stackrel{\text{SK}}{\longleftrightarrow}\text{DO}) $, indicating that the DU believes that the DO believes in the shared key (SK) between DU and DO.

 Goal $G_2$: $\text{DU}|\equiv (\text{DU}\stackrel{\text{SK}}{\longleftrightarrow}\text{DO}) $, indicating that the DU believes in the shared key (SK) between DU and DO.

 Goal $G_3$: $\text{DO}|\equiv \text{DU}|\equiv (\text{DO}\stackrel{\text{SK}}{\longleftrightarrow}\text{DU}) $, indicating that DO believes that DU believes in the shared key (SK) between DO and DU.

 Goal $G_4$: $\text{DO}|\equiv (\text{DO}\stackrel{\text{SK}}{\longleftrightarrow}\text{DU}) $, indicating that DO believes in the shared key (SK) between DO and DU.

 \subsubsection{Formulating Assumptions}
 Before analyzing the protocol, we must establish the initial belief assumptions necessary for the protocol's completion. These assumptions are the conditions required for each message in the protocol to function correctly. Based on this scheme, we propose the following initial assumptions.

Assumption $a_1$: $\text{DU}|\equiv \#(T_1),\text{DU}|\equiv \#(T_4),\text{DU}|\equiv \#(T_5)$

Assumption $a_2$: $\text{DO}|\equiv \#(T_1),\text{DO}|\equiv \#(T_2),\text{DO}|\equiv \#(T_5)$

Assumption $a_3$: ${\text{Hospital}_\text{A}}|\equiv \#(T_2),{\text{Hospital}_\text{A}}|\equiv \#(T_3)$

Assumption $a_4$: $\text{RelyChain}|\equiv \#(T_3),\text{RelyChain}|\equiv \#(T_4)$

Assumption $a_5$: $\text{DU}|\equiv \xrightarrow[]{pk_{DO}}\text{DO}$

Assumption $a_6$: $\text{DO}|\equiv \xrightarrow[]{pk_{DU}}\text{DU}$

Assumption $a_7$: $\text{DU}|\equiv \text{DO}| \Rightarrow \text{DU}\stackrel{\text{SK}}{\longleftrightarrow}\text{DO}$

Assumption $a_8$: $\text{DO}|\equiv \text{DU}| \Rightarrow \text{DO}\stackrel{\text{SK}}{\longleftrightarrow}\text{DU}$

\subsubsection{Analyzing the Protocol}
In this part, we ensure the correctness and security of the \schemeName scheme by using the messages from the idealized protocol, the initial assumptions, and the logical inference rules of BAN logic to achieve the desired goals of the protocol.

It can be inferred from the message $M_3$:
\vspace{-1ex}
\begin{equation}
\begin{aligned}
    \text{DO}\triangleleft \{ request_1,pk_{DO},pk_{DU},T_1,N_1,\text{DO}\stackrel{\text{SK}}{\longleftrightarrow}\text{DU}\}_{pk_{DO}}
\label{PA1}
\end{aligned}
\end{equation}

Based on the assumption $a_6$, formula (\ref{PA1}), we can conclude that:
\begin{equation}
\begin{aligned}
\text{DO}|\equiv \text{DU}|\sim\{&request_1,pk_{DO},pk_{DU},T_{1},N_1,\\&\text{DO}\stackrel{\text{SK}}{\longleftrightarrow}\text{DU}\}_{pk_{DO}}
\label{PA2}
\end{aligned}
\end{equation}

From the assumption $a_1$, we can conclude that:
\begin{equation}
\begin{aligned}
\text{DO}|\equiv \#(\{&request_1,pk_{DO},pk_{DU},T_1,N_1,\\&\text{DO}\stackrel{\text{SK}}{\longleftrightarrow}\text{DU}\}_{pk_{DO}})
\label{PA3}
\end{aligned}
\end{equation}

From formula (\ref{PA2}), (\ref{PA3}), we can conclude that:
\begin{equation}
\begin{aligned}
   \text{DO}|\equiv \text{DU}|\equiv(\{ &request_1,pk_{DO},pk_{DU},T_1,N_1,\\&\text{DO}\stackrel{\text{SK}}{\longleftrightarrow}\text{DU}\}_{pk_{DO}})
\label{PA4}
\end{aligned}
\end{equation}

From the formula (\ref{PA4}), we can conclude that:
\begin{equation}
  \text{DO}|\equiv \text{DU}|\equiv (\text{DO}\stackrel{\text{SK}}{\longleftrightarrow}\text{DU})
\label{PA5}
\end{equation}

Formula (\ref{PA5}) meets the goal $G_3$.

Based on the assumption $a_8$, formula (\ref{PA5}), we can conclude that:
\begin{equation}
   \text{DO}|\equiv(\text{DO}\stackrel{\text{SK}}{\longleftrightarrow}\text{DU})
\label{PA6}
\end{equation}

Formula (\ref{PA6}) meets the goal $G_4$.

It can be inferred from the message $M_5$:
\begin{equation}
\begin{aligned}
       \text{DU}\triangleleft \{&respond_2,,pk_{DO},pk_{DU},T_5,N_4,\\&\text{DO}\stackrel{\text{SK}}{\longleftrightarrow}\text{DU}\}_{pk_{DU}}
\label{PA7}
\end{aligned}
\end{equation}

From assumption $a_5$, formula (\ref{PA7}), we can conclude that:
\begin{equation}
\begin{aligned}
\text{DU}|\equiv \text{DO}|\sim\{&respond_2,pk_{DO},pk_{DU},T_5,N_4,\\&\text{DO}\stackrel{\text{SK}}{\longleftrightarrow}\text{DU}\}_{pk_{DU}})
\label{PA8}
\end{aligned}
\end{equation}

From the assumption $a_2$, we can conclude that:
\begin{equation}
\begin{aligned}
\text{DU}|\equiv \#(\{&respond_2,pk_{DO},pk_{DU},T_5,N_4,\\&\text{DO}\stackrel{\text{SK}}{\longleftrightarrow}\text{DU}\}_{pk_{DU}})
\label{PA9}
\end{aligned}
\end{equation}

From formula (\ref{PA8}), (\ref{PA9}), we can conclude that:
\begin{equation}
\begin{aligned}
\text{DU}|\equiv \text{DO}|\equiv (\{&respond_2,pk_{DO},pk_{DU},T_5,N_4,\\&\text{DO}\stackrel{\text{SK}}{\longleftrightarrow}\text{DU}\}_{pk_{DU}} \})
\label{PA10}
\end{aligned}
\end{equation}

From the formula (\ref{PA10}), we can deduce that:
\begin{equation}
\text{DU}|\equiv \text{DO}|\equiv(\text{DO}\stackrel{\text{SK}}{\longleftrightarrow}\text{DU})
\label{PA11}
\end{equation}

Formula (\ref{PA11}) meets goal $G_1$.

From the assumption $a_7$, formula (\ref{PA11}), it can be concluded that:
\begin{equation}
     \text{DU}|\equiv (\text{DO}\stackrel{\text{SK}}{\longleftrightarrow}\text{DU})
\label{PA12}
\end{equation}

Formula (\ref{PA12}) meets goal $G_2$.

To sum up, formula (\ref{PA5}), formula (\ref{PA6}), formula (\ref{PA11}) and formula (\ref{PA12}) meet the four expected goals set by the protocol, proving its logical correctness. This demonstrates the rationality of the protocol and shows that the \schemeName scheme ensures both data confidentiality and integrity.

\subsection{Security model}
\label{sec:5.5.4}
The algorithm in MedExChain is defined as follows: System Setup stage (\ref{1.1}) is defined as $Setup$. In Key Generation (\ref{2.1}), (2.1) phase is defined as  \textit{KeyGen}, and (2.2) phase is defined as \textit{CRF-KeyGen}. In Data Encryption (\ref{3.1}), (3.1) phase is defined as \textit{Enc}, and (3.2) phase is defined as \textit{CRF-Enc}. In the Data Request (\ref{4.2}), phases (4.2), (4.3) and (4.5) are defined as \textit{ReKeyGen, CRF-ReKeyGen} and  \textit{ReEnc} respectively. We reviewed relevant literature addressing analogous issues and formalized two security models as a result \cite{71}, \cite{12}.
\subsubsection{Chosen-Plaintext Attack (CPA) security model}The game is played by challenger $\mathcal{C}$ and adversary $\mathcal{A}$.
\begin{enumerate}
\item [(1)] 
\textbf {Initial:} $\mathcal{C}$ runs the $Setup$ algorithm to generate the public parameter  and sends it to $\mathcal{A}$.
\item [(2)] 
\textbf {Phase 1\&2:} $\mathcal{A}$ can perform polynomial bounded number of queries.\par 
\textbullet  \quad 
Key generation Oracle $O_{sk}$: Given  and identity ID; $\mathcal{C}$ runs  \textit{KeyGen} algorithm to generate private key $sk_{ID}$ and send $sk_{ID}$ to CRF. CRF runs  \textit{CRF-KeyGen} algorithm to obtain the re-randomized user private key $sk_{ID}^\prime$, and then sends it to $\mathcal{A}$. Let $ \Gamma_U$ be the user index set.\par

\textbullet  \quad 
Cryptographic Oracle $O_{en}$: When the adversary inputs ID and access structure A, the $\mathcal{C}$ first calculates the ciphertext $C_{ID}$ through  \textit{Enc} algorithm, and the re-randomized ciphertext $C_{ID}^\prime$ is obtained by running  \textit{CRF-Enc} algorithm, and then it is sent to $\mathcal{A}$.\par

\textbullet  \quad 
Re-encryption key generation Oracle $O_{rk}$: When the adversary inputs $(ID_i,ID_j)$ , where $i\in \Gamma_U$. The re-encryption key calculated by  \textit{ReKeyGen} and  \textit{CRF-ReKeyGen} corresponds to $(ID_i,ID_j)$, and then the key is output. If $i=j$, $\mathcal{C}$ quits the game.\par

\textbullet  \quad 
Re-encryption Oracle $O_{re}$: When the adversary inputs $(pk_i,pk_j,C_i)$, where  $pk_i,pk_j$ comes from $O_{sk}$. $\mathcal{C}$ first runs  \textit{ReKeyGen} and  \textit{CRF-ReKeyGen} algorithms to get the re-encrypted key $rk_{i\rightarrow j}^\prime$, and then runs ReEnc algorithm to calculate the re-encrypted ciphertext $C_j$ according to $C_i$, and sends the $C_j$ to $\mathcal{A}$. Among them, $C_j$ can be decrypted by $sk_j$. If $i=j$, $\mathcal{C}$ quit the game.\par

\item [(3)] 
\textbf {Challenge:} $\mathcal{A}$ decides when Phase 1 ends, and then generates two messages $m_0$ and $m_1$ of equal length, which wants to be challenged. $\mathcal{C}$ takes a random bit $b\in \{0,1\}$, calculates the ciphertext $C_i= \textit{Enc}(par,pk_i,m_b)$, and then runs \textit{CRF-Enc} algorithm to get the re-randomized ciphertext $C_i^\prime$, which is sent to $\mathcal{A}$ as the questioned ciphertext.\par 

\item [(4)] 
\textbf {Guess:}
If $b^\prime=b$, $\mathcal{A}$ outputs bit $b^\prime$ and wins the game.\par
We define the advantage of $\mathcal{A}$ attacking this scheme as: $Adv_\mathcal{A}^{CPA}=|Pr[b^\prime=b]-1/2|$, and $Pr[b^\prime=b]$ represents the probability of $b^\prime=b$.
\end{enumerate}
\par 
\begin{definition}
\textbf {CPA security for MedExChain.} If there is no adversary $\mathcal{A}$ with bounded polynomials who has an advantage over the challenger $\mathcal{C}$ in the game, then MedExChain scheme is $(\epsilon,t,q_{sk},q_{en},q_{rk})$-$CPA$ security. Among them, $\epsilon$ is the advantage of $\mathcal{A}$ in winning the game, $T$ is the running time of the game, $q_{sk}$ is the number of key generation queries, $q_{en}$ is the number of encrypted queries, and $q_{rk}$ is the number of re-encrypted key generation queries.
\end{definition}

\subsubsection{Algorithm Substitution Attack (ASA) security model}The adversary $\mathcal{A}$ can replace any algorithm except the algorithm of CRF operation, and then attack the system. What is special about ASA is that the algorithm is manipulated unconsciously. Thereby causing the disclosure of user's secret information. MedExChain scheme can achieve exfiltration-resistant security through ASA proof. The game is played by the challenger $\mathcal{C}$ and the adversary $\mathcal{A}_t$, as shown below.
\begin{enumerate}
\item [(1)]  
\textbf {Tempering:}
$\mathcal{A}$ selects some tampered algorithms $Setup*$, $KeyGen*$, $ReKeyGen*$, $Enc*$, and then sends them to $\mathcal{C}$. $\mathcal{C}$ replaces its original algorithm with these tampered algorithms after receiving them.
\par 

\item [(2)]  
\textbf {Initial:} $\mathcal{C}$ runs  \textit{Setup*} algorithm and  \textit{KeyGen*} algorithm to generate public parameter $par$ and key pair $(pk_{ID},sk_{ID})$, and sends $pk_{ID}$ and $par$ to $\mathcal{A}$, and keep secret $pk_{ID}$ to $\mathcal{A}$.
\par

\item [(3)]  
\textbf {Phase 1\&2:} Same as Phase 1\&2 in Sec \ref{3.1}, except that $\mathcal{C}$ uses  \textit{ReKeyGen*} and  \textit{Enc*} instead of  \textit{ReKeyGen} and  \textit{Enc}.
\par 

\item [(4)]  
\textbf {Challenge:} $\mathcal{A}$ decides when the first phase will end. $\mathcal{A}$ generates two messages or public keys of equal length, which it wants to challenge. $\mathcal{C}$ takes a random bit $b\in \{0,1\}$, and then calculates the challenged re-random ciphertext or the challenged private key of $\mathcal{A}$. If the ciphertext or key of the target message $b^*$ or key $pk^*$ fails, then $\mathcal{A}$ fails in this game.
\par

\item [(5)]  
\textbf {Guess:} If $b^\prime=b$, $\mathcal{A}$ outputs bit $b^\prime$ and wins the game.\par
We define the advantage of $\mathcal{A}$ attacking this scheme as: $Adv_\mathcal{A}^{exf-res}=|Pr[b^\prime=b]-1/2|$, and $Pr[b^\prime=b]$ represents the probability of $b^\prime=b$.
\par

\end{enumerate}
\begin{definition}
\textbf {Weak exfiltration-resistant for CRF.} If there is no adversary $\mathcal{A}$ with polynomial boundedness who has a non-negligible advantage over the challenger $\mathcal{C}$ in the game, then the MedExChain scheme is weakly $(\epsilon,t,q_{sk},q_{rk},q_{re})$-exfiltration-resistant security. Among them, $\epsilon$ is the advantage of $\mathcal{A}$ to win the competition, $t$ is the running time of the competition, $q_{sk}$ is the number of key generation queries, $q_{rk}$ is the number of re-encryption key generation queries, and $q_{re}$ is the number of re-encryption queries. Game is the antagonistic implementation of functional maintenance, which means that the exfiltration-resistant ability is weak.
\end{definition}

\subsection{Security Analysis}
\label{sec:5.5.5}

\textbf{Theorem 1.}
The CRFs of Hospital and Data Owner in MedExChain scheme are weak security-preserving and exfiltration-resistant. Among them, weak security-preserving means that Ateniese's IBPRE scheme \cite{20} is CPA security. Exfiltration-resistant means that MedExChain scheme can resist information leakage when faced with an adversary who initiates ASA that does not affect normal functions.

\subsubsection{{Weak security-preserving}}
We use the tampering algorithms  \textit{KeyGen*} and  \textit{ReKeyGen*} to prove the CPA security of MedExChain scheme, and \textit{Enc*} proves the indistinguishability between MedExChain and IBPRE's security game. First of all, we have introduced the CPA security model in the previous section, and the standard security game is introduced in IBPRE. Next, let's consider the following game.

\begin{enumerate}
\item [(1)]  
\textbf{Game 0.} Similar to the CPA security model in Sec \ref{3.1}.
\par

\item [(2)]  
\textbf{Game 1.} 
Same as Game 0, except that the user's private key is generated by \textit{KeyGen} in the standard security game, instead of \textit{KeyGen*} and \textit{CRF-KeyGen} algorithms in Phase 1\&2.
\par

\item [(3)]  
\textbf{Game 2.} 
Same as Game 1, except that the re-encryption key $rk_i$ is generated by \textit{ReKeyGen} in the standard security game, instead of \textit{ReKeyGen*} and \textit{CRF-ReKeyGen} algorithms during Phase 1\&2.
\par

\item [(4)]  
\textbf{Game 3.} 
Same as Game 1, except that the ciphertext $\mathcal{C}_i$ is generated by \textit{Enc} in the standard security game, instead of \textit{Enc*} and \textit{CRF-Enc}  algorithms of Phase 1\&2.
\par

\item [(5)]  
\textbf{Game 4.} 
Same as Game 3, except that the challenged ciphertext $\mathcal{C}_i^\prime$ is generated by \textit{Enc} in the standard security game, instead of \textit{Enc*} and \textit{CRF-Enc} algorithms during the challenge.
\par

\end{enumerate}

Then we prove the inseparability between Game 0 and Game 1, Game 1 and Game 2, Game 2 and Game 3, and Game 3 and Game 4 respectively.

\begin{enumerate}
\item[(1)]
\textbf{Game 0 and Game 1.} Suppose there is a tampered algorithm \textit{KeyGen*}, after running \textit{CRF-KeyGen} algorithm processed by CRF, the updated user private key $sk_i^\prime$ is generated. It is a consistent random number, because the user's private key $sk_i$ has key extensibility, which is the same as the original algorithm \textit{KeyGen}. Therefore, Game 0 and Game 1 are indistinguishable.
\par

\item[(2)]
\textbf{Game 1 and Game 2.} Suppose there is a tampered algorithm \textit{ReKeyGen*}, and after running \textit{CRF-ReKeyGen} algorithm processed by CRF, an updated re-encryption key $rk_i^\prime$ is generated. It is a consistent random number because the re-encryption key $rk_i$ has key extensibility, which is the same as the original algorithm \textit{ReKeyGen}. Therefore, Game 1 and Game 2 are indistinguishable.
\par

\item[(3)]
\textbf{Game 2 and Game 3.} Suppose there is a tampered algorithm Enc*, and after running \textit{CRF-Enc}, the post-processed encrypted text $\mathcal{C}_i^\prime$ is generated. It is a consistent random number, because the IBPRE scheme can be re-randomized, which is the same as the Enc algorithm in IBPRE, and has nothing to do with the behavior of Enc*. Therefore, Game 2 and Game 3 are indistinguishable.
\par

\item[(4)]
\textbf{Game 3 and Game 4.} For the same reason as Game 2 and Game 3.
\end{enumerate}
\par
Because the IBPRE scheme is CPA security, the tampered MedExChain scheme can also achieve CPA security. Therefore, we can conclude that the CPA security of MedExChain indicates that the CRF of Hospital and data owner is weak security-preserving.

\subsubsection{Exfiltration-resistant}
The indistinguishability between Game 0 and Game 4 indicates that CRF of Hospital and data owner is weak security-preserving.

\subsection{Formal verification based on Scyther tool}
\label{sec:5.2}

\begin{table}[h]
\footnotesize
\caption{\MakeUppercase {The modeling and implementation process in \emph{Scyther}}}
\label{Implementation}
\centering
\begin{tabular}{|p{4.0cm}|p{4.0cm}|}
\hline
 \textbf{Protocol Step} &  \textbf{Modeling Implementation}\\ 
    \hline
{1. DU generates a request message.} &  {fresh T1: Timestamp;
fresh request, key: text;} \\ 
\hline
 2. DU transmits a request message to DO.&
$\mathsf{(DU, DO,}$\par
$\mathsf{\{DU, DO, request, T1\}pk(DO));}$\\
\hline
 3. The DO receives the request message from the DU.&  {$\mathsf{recv\_1(DU, DO,}$\par
 $\mathsf{\{DU, DO, request, T1\}pk(DO) );}$}
\\
\hline
4. DO perform the verification operation.
& {$\mathsf{fresh T2: Timestamp;}$\par
($\mathsf{fresh Data1, RK: text;}$\par
$\mathsf{var T1: Timestamp;}$\par
$\mathsf{var request, ACK1, response: text;}$\par
$\mathsf{match(ACK1, H(request));}$}\\
\hline
5. DO transmits a communication to Hospital A.&$\mathsf{send\_2(DO, HospitalA, \{DO,}$\par
$\mathsf{HospitalA, Data1, RK\}}$\par
$\mathsf{pk(HospitalA));}$\\
\hline
6. Hospital A receives a message from DO.& 
 $\mathsf{recv\_2(DO, HospitalA,}$\par
 $\mathsf{{DO, HospitalA, Data1, RK}}$\par
 $\mathsf{pk(HospitalA));}$
\\
\hline
7. Hospital A acquires the encrypted data ciphertext M1.
&
$\mathsf{fresh T3: Timestamp;}$\par
$\mathsf{var M1, Data1, RK: text;}$\par
$\mathsf{match(M1, H(Data1));}$
\\
\hline
8. Hospital A transmits a message to the Relay.&   $\mathsf{send_3(HospitalA,Relay,}$\par
$\mathsf{\{HospitalA, Relay, M1, RK\}}$\par
$\mathsf{pk(Relay));}$\\
\hline
9. Relay receives the message from Hospital A.& {$\mathsf{recv\_3(HospitalA, Relay,}$\par
$\mathsf{\{HospitalA, Relay, M1, RK\}}$\par
$\mathsf{pk(Relay));}$}\\
\hline
10. Relay produces M2 and Data2.
&
$\mathsf{fresh T4: Timestamp;}$\par
$\mathsf{fresh M2: text;}$\par
$\mathsf{var M1, RK, Data2: text;}$\par
$\mathsf{match(M2, H(M2, RK));}$\par
$\mathsf{match(Data2, H(M2, RK));}$
\\
\hline
11. Relay transmits Data2 to Hospital A.
& {$\mathsf{send\_4(Relay, HospitalA,}$\par
$\mathsf{\{Relay, HospitalA, Data2\}}$\par
$\mathsf{pk(HospitalA));}$}\\
\hline
12. Hospital A receives Data2.
& {$\mathsf{recv\_4(Relay, HospitalA,}$\par
$\mathsf{\{Relay, HospitalA, Data2\}}$\par
$\mathsf{pk(HospitalA));}$}\\
\hline
13. Hospital A initiates a response.
& {$\mathsf{var Data2, response: text;}$\par
$\mathsf{match(response, H(Data2));}$
}\\
\hline
14. HospitalA submits a response to DO.
& {$\mathsf{send\_5(HospitalA, DO,}$\par
$\mathsf{\{HospitalA, DO, response\} }$\par
$\mathsf{pk(DO));}$
}\\
\hline
15. The DO receives a response.
& {$\mathsf{recv\_5(HospitalA, DO,}$\par
$\mathsf{\{HospitalA, DO, response\}}$\par
$\mathsf{pk(DO));}$
}\\
\hline
16. DO sends response to DU.
& {$\mathsf{send\_6(DO, DU,}$\par
$\mathsf{\{DO, DU, response\}pk(DU));}$
}\\
\hline
17. DU receives DO's response.
& {$\mathsf{recv\_6(DO, DU,}$\par
$\mathsf{\{DO, DU, response\}pk(DU));}$
}\\
\hline
18. Relay transmits M2 to DU.
& {$\mathsf{send\_7(Relay, DU,}$\par
$\mathsf{{Relay, DU, M2}pk(DU));}$
}\\
\hline
19. DU receives M2 from the Relay.
& {$\mathsf{recv\_7(Relay, DU,}$\par
$\mathsf{\{Relay, DU, M2\}pk(DU));}$
}\\
\hline
20. DU decrypts M2 using the key.
& {$\mathsf{fresh key: text;}$\par
$\mathsf{var response, M2, m: text;}$\par
$\mathsf{match(m,(response, M2, key))}$}\\
\hline
\end{tabular}

\end{table}
\par
\emph{Scyther} \cite{26}, a tool for verifying protocol security attributes, was initially developed to deeply analyze and verify critical security elements such as confidentiality, authentication, and data integrity. In the context of authentication, \emph{Scyther} is integrated with the Security Protocol Description Language (SPDL) to specify the protocol and verify the security of predefined assumptions. The \emph{Scyther} tool can effectively identify potential security issues within the protocol based on these assumptions. It offers four types of statements: Aliveness, Weak agreement, Non-injective agreement, and Non-injective synchronisation. These statements, in conjunction with the SPDL, provide a comprehensive and multi-dimensional approach to protocol security analysis, ensuring the thoroughness and accuracy of the verification process.



\subsubsection{Formal Protocol Modeling}
In this framework, we delineate four distinct roles: Data Owner (DO), Data User (DU), Hospital A, and the Relay Chain.
\begin{enumerate}
\item [(1)] Entity Declaration
   
$usertype\ text$; \par
$usertype\ Timestamp$; \par 
$hashfunction\ H$; \par

Among these, `hashfunction' represents a built-in hash function, while `usertype' denotes a user-defined type.
\item [(2)] Modeling and Implementation\par 
The modeling and implementation process is shown in Table~\ref{Implementation}. 
\end{enumerate}


\subsubsection{Formal Security Attribute Modeling} 
\par
The \emph{Scyther} formal verification tool does not directly verify authentication but requires the verification of authentication through attributes such as Secret, Alive, Weakagree, Niagree, and Nisynch. The table below provides an example to illustrate the modeling of security attributes in the formal verification of protocols.
\par
Among these, the strength of authentication attributes—Aliveness, Weak agreement, Non-injective agreement, and non-injective synchronisation—increases progressively. Aliveness is a fundamental attribute, ensuring the presence of the anticipated communication party A. Weakagree denotes weak agreement authentication, requiring that certain states or values among participants remain consistent throughout the agreement's execution. Non-injective agreement (Niargree) is a non-monotonic consistency authentication, describing that the communication or negotiation outcomes among participants cannot be repudiated during the protocol's execution. Non-injective synchronisation (Nisynch) signifies that all send/receive events preceding a claim event can be executed by the correct agent, A, in the correct order and content, even when the attacker possesses A's private key. This property ensures the integrity of the information received by the receiver and describes that the communication or negotiation results among participants cannot be denied and remain consistent. Although Nisynch and Niargree are conceptually similar, Nisynch imposes stricter requirements on the expected order, thereby offering stronger authentication.\par

\par
\subsubsection{Analysis of Formal Verification Outcomes} 
In the \emph{Scyther} model, this paper introduces four distinct roles: DO, DU, HospitalA, and Relay. When any two of these entities exchange information, a timestamp is employed to ensure the message's timeliness. The protocol is described using SPDL. Initially, DU sends a cross-chain access request message to DO, which, upon receipt, verifies the request. Following verification, DO calculates the key RK and transmits the locally stored Data1 and PK to HospitalA. HospitalA then derives the ciphertext M1 from Data1 and forwards M1 and RK to Relay. Upon receiving M1 and RK, Relay generates the re-encrypted ciphertext M2 and the ciphertext identifier Data2, sending Data2 back to HospitalA. HospitalA subsequently sends a response message containing Data2 to DO, which in turn relays the response to DU. DU decrypts M2 received from Relay using its private key to obtain the relevant data M. DU decrypts M2 received from Relay using its private key to obtain the associated data m.By running the SPDL model of this protocol and testing the security assertions of the four parties, it is demonstrated that the protocol can effectively resist replay attacks, man-in-the-middle attacks, and other potential threats. In summary, the \emph{Scyther} tool's proof confirms that no suspicious security attacks exist within this protocol. The verification results are illustrated in Figure \ref{fig5}.
\begin{figure}[htbp]
\centering
\includegraphics[width=0.48\textwidth]{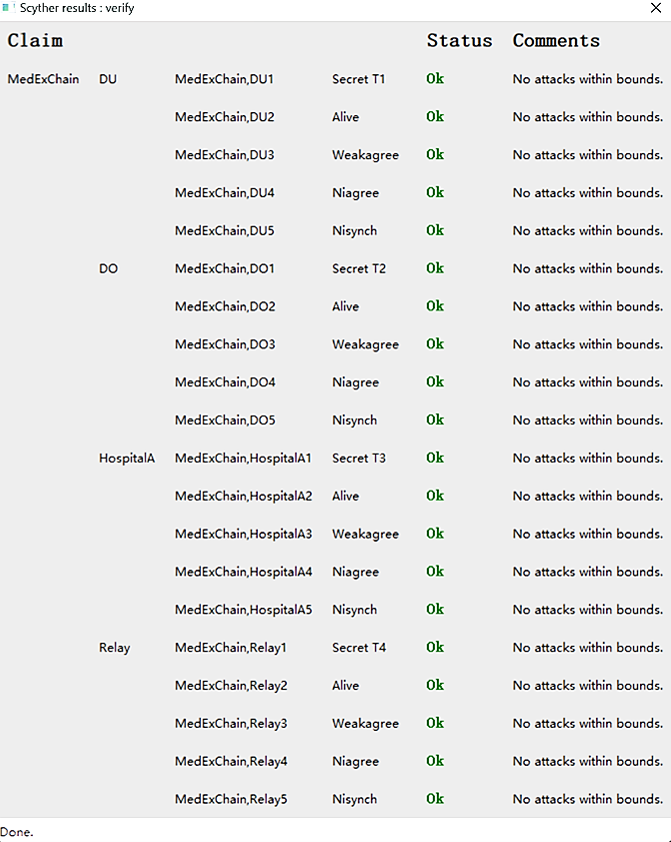}
\caption{\emph{Scyther} tool verification results for MedExChain}
\label{fig5}
\end{figure}
{\section{Performance Analysis}\label{PA}}
\begin{table*}[htbp]
\centering
\renewcommand{\arraystretch}{1.3}
\begin{threeparttable}[c]
\caption{{\MakeUppercase {Comparison of Computational Overhead}}}
\label{table:computationalOverhead}
\begin{tabular}{|l|c|c|c|c|c|c|c|}
\hline
\bfseries Scheme & \bfseries \textbf{$KeyGen_{DO}$} & \bfseries \textbf{$KeyGen_{DU}$} & \bfseries \textbf{$Enc$} & \bfseries \textbf{$ReKeyGen$} & \bfseries \textbf{$ReEnc$} & \bfseries \textbf{$Dec$} & \bfseries Total (ms)\\
\hline
\bfseries IBPRE\_CRF \cite{12} & 2$E1$+$H$ & 2$E1$+$H$ & 4$E1$+2$E2$+3$P$ & 3$E1$+2$E2$+2$P$+$H$ & $P$ & 2$P$+$H$ & 96.02\\
\hline
\bfseries CDSS \cite{21} & 6$E1$+2$H$ & 7$E1$ & 4$E1$+2$E2$+3$H$+$P$ & 5$E1$+$E2$+$P$+2$H$ & $E1$+$2P$+$P$ & $E2$ & 154.05\\
\hline
\bfseries ABE-IBE \cite{17} & (2$N$+1)$E1$ & 4$E1$+$H$ & ($N$+1)$E1$+$E2$ & (4$N$+4)$E1$+$E2$+$P$ & \makecell{2$E1$+$E2$+\\(2$N$+1)$P$} & $2P$ & 322.77\\
\hline
\bfseries CP-HAPRE \cite{18} & (5$n$+7)$E1$ & (5$n$+7)$E1$ & (6$N$+1)$E1$+2$E2$+$P$ & 5$E1$+2$E2$+$P$+$H$ & ($n$+1)$E2$+3$nP$ & 3$P$+$H$ & 491.36\\
\hline
\bfseries FABRIC \cite{27} & 15$E1$+6$H$ & \makecell{(9$n$+15)$E1$\\+(6$n$+6)$H$} & 12$E1$+2$E2$+6$H$ & \makecell{($9N^2-3N+9$)\\ $E1$+2$E2$+(6$N^2+1) H$} & 6$P$ & \makecell{6$nE1$+\\9$P$+$H$} & 1,856.46 \\
\hline
\bfseries \schemeName (Ours) & 2$E1$+$H$ & 4$E1$+2$H$ & 3$E1$+2$E2$+2$P$ & 3$E1$+2$E2$+2$P$+$H$ & $P$ & 2$P$+$H$ & 98.24\\
\hline
\end{tabular}
\begin{tablenotes}
\centering
\item $E1$: An exponentiation over group $G_1$ (5,689$\mu$s)\quad $E2$: An exponentiation over group $G_2$ (498$\mu$s)\quad
$H$: Hash function operation (269$\mu$s) \\$P$: The operation of bilinear pairing (3,823$\mu$s)\quad
$n$: The number of attributes required by the access policy in ABE\\
$N$: The total number of attributes included in the access policy in ABE\\
\end{tablenotes}
\end{threeparttable}
\end{table*}

In this section, we assess the computational and communication overhead, as well as the performance of the blockchain system. The experiments were conducted on a laptop equipped with an AMD Ryzen 7 5800H processor with Radeon graphics, clocked at 3.20 GHz, 16.0 GB of RAM, running Ubuntu 22.04.2 as the operating system. The programming language used was Java 1.8, with bilinear pairing operations facilitated through the JPBC library. The elliptic curve employed for constructing the bilinear pairing is of Type A, with a system security factor of 80 bits. The blockchain platform utilized is FISCO-BCOS v3.6.0. \par
We compare our MedExChain with IBPRE\_CRF \cite{12}, CDSS \cite{21}, ABE-IBE \cite{17}, CP-HAPRE \cite{18} and FABRIC \cite{27} because these schemes, similar to ours, implement encryption system conversion via proxy re-encryption. Notably,\cite{17} was the first to introduce the concept of encryption system conversion. It is important to mention that the schemes \cite{17}, \cite{18}, \cite{27} incorporate attribute-based encryption, which inherently has disadvantages compared to identity-based encryption. To ensure a fair comparison, we adopt the simplest access policy for attribute-based encryption in our analysis. The access policy includes a total of 5 attributes (n), with users possessing 3 attributes (n).\par
Given the design advantages of the MedExChain scheme, with the exception of the PHR $Enc$ stage, the primary operations are executed external to the IoMT devices. Our evaluation not only assesses whether the $Enc$ stage meets the performance requirements of IoMT devices but also benchmarks the performance of each stage against existing schemes.\par
 \vspace{\baselineskip}
\subsection{Computational Overhead}
\subsubsection{Theoretical Analysis} We conducted a theoretical analysis to calculate the computational overhead of core operations in each scheme, including the key pair generation operations for the data owner ($KeyGen_{DO}$) and data user ($KeyGen_{DU}$), ciphertext generation ($Enc$), re-encryption key generation ($ReKeyGen$), re-encryption ciphertext generation ($ReEnc$), and decryption ($Dec$). We focused on operations with significant computational overhead, such as bilinear pairing ($P$), exponentiation in the $G1$ group ($E1$), exponentiation in the $G2$ group ($E2$), and hash function ($H$). Operations involving addition, subtraction, multiplication, and division were disregarded due to their relatively low computational overhead.\par
The MedExChain scheme not only enhances system security but also maintains a computational overhead advantage. The results of our theoretical analysis are detailed in Table \ref{table:computationalOverhead}. By comparing the coefficients of each operation, it is evident that the MedExChain scheme is advantageous in terms of computational overhead across all stages, particularly at the $Enc$, $ReKeyGen$, and $ReEnc$ stages. At the $Enc$ stage, the MedExChain scheme requires only three $E1$ operations, two $E2$ operations, and two $P$ operations to generate ciphertext, which is less than all other schemes. During the $ReKeyGen$ stage, the MedExChain scheme requires three $E1$ operations, two $E2$ operations, two $P$ operations, and one $H$ operation to generate the re-encryption key, matching the \cite{12} scheme and outperforming others. At the $ReEnc$ stage, the MedExChain scheme completes the generation of re-encrypted ciphertext with just one $P$ operation, equaling the \cite{12} scheme and significantly outperforming other schemes. These three stages are crucial for data sharing among users. The computational overhead advantage of the MedExChain scheme in these stages conserves the computing power of IoMT devices, enabling them to handle more data sharing requests more efficiently.\par
In our experimental environment, we measured the time required for each operation: $P$: 3823 $\mu$s, $E1$: 5689 $\mu$s, $E2$: 498 $\mu$s, and $H$: 269 $\mu$s. We then calculated the theoretical total computational overhead for all stages of each scheme. As shown in the Total (ms) column of Table \ref{table:computationalOverhead}, the MedExChain scheme is equivalent to the \cite{12} scheme and significantly lower than other schemes.\par

\begin{figure*}[!t]
    \centering
    \includegraphics[width=17.6cm]{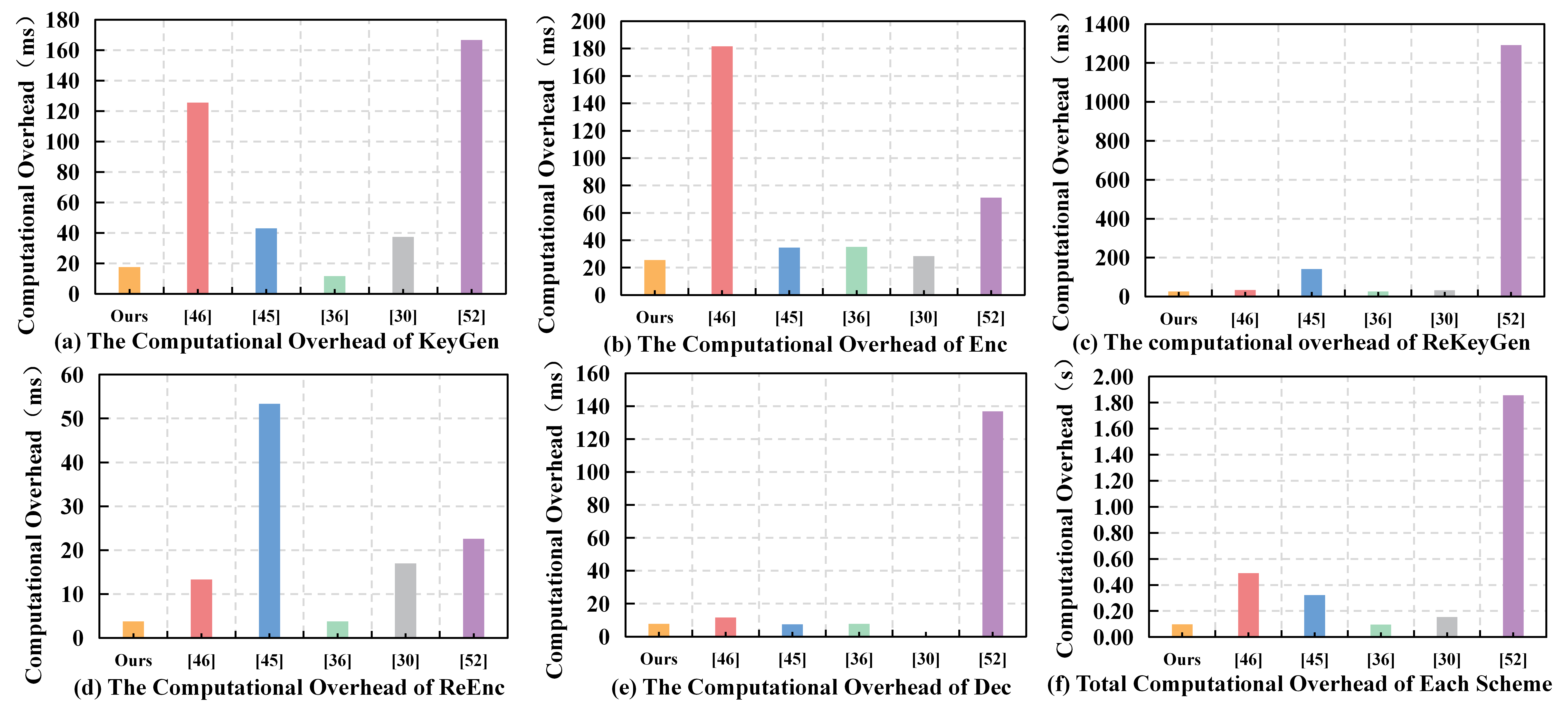}
    \caption{Comparison of Computational Overhead}
    \label{fig:11}
\end{figure*}

\begin{figure*}[!t]
    \centering
    \includegraphics[width=17.6cm]{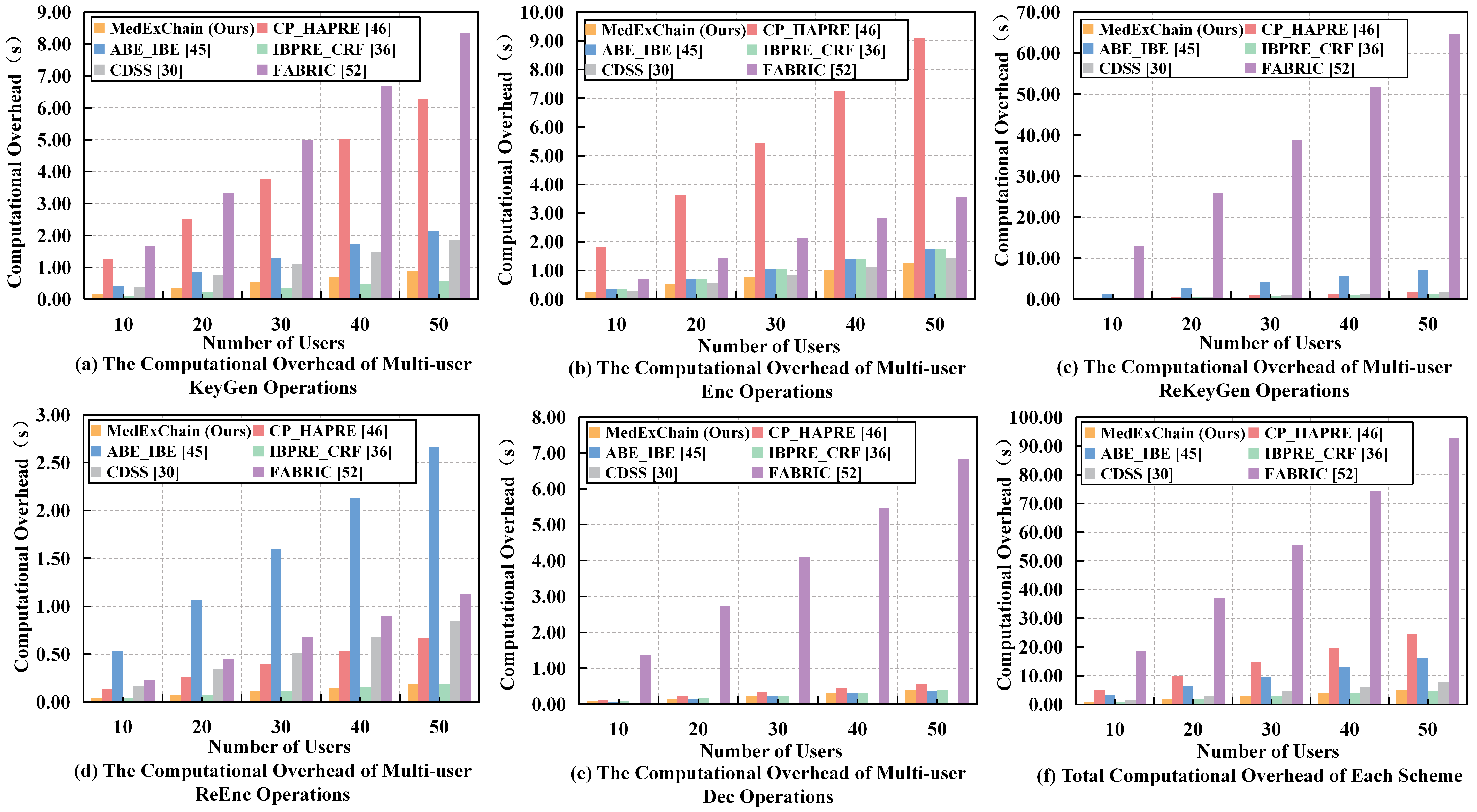}
    \caption{The Comparison of Computational Overhead for Multi-user Access}
    \label{fig:Computational_Overhead1}
\end{figure*}

\subsubsection{Experimental Measurement} In the aforementioned experimental setup, we implemented the code for each scheme and measured the execution time required for each stage. Specifically, we executed each stage 50 times and calculated the average execution time. The experimental results, depicted in Figure \ref{fig:11}, align with our theoretical analysis. The MedExChain scheme demonstrates a favorable position in terms of the execution time required for each stage and the total time to complete all stages. Figure \ref{fig:11}(a) illustrates the execution time of the $KeyGen$ stage for each scheme. For comparative purposes, we computed the average execution time of the $KeyGen_DO$ and $KeyGen_DU$ stages for each scheme. The execution time of the MedExChain scheme is slightly longer than that of the \cite{12} scheme but significantly shorter than that of other schemes. This is primarily because the MedExChain scheme includes two additional E1 operations in the $KeyGen_DU$ stage compared to the \cite{12} scheme, while other schemes involve more E1 operations, which are notably time-consuming. Figure \ref{fig:11}(b) presents the execution time of the $Enc$ stage for each scheme. The MedExChain scheme's execution time is slightly shorter than that of the \cite{17}, \cite{12}, and \cite{21} schemes and significantly shorter than that of the other two schemes. This is mainly due to the $E1$ operation in the \cite{17}, \cite{12}, \cite{21} schemes being slightly longer than that in the MedExChain scheme. Figure \ref{fig:11}(c) shows the execution time of the $ReKeyGen$ stage for each scheme. The MedExChain scheme's execution time is equivalent to that of the \cite{18}, \cite{12}, and \cite{21} schemes but considerably less than that of the \cite{17} and \cite{27} schemes. This is because the time complexity of the $ReKeyGen$ stage in the \cite{17}, \cite{27} schemes is correlated with the number of attributes in the access policy, with a significant correlation coefficient. Even with a simple access policy set in the experiment, the execution time required for the two schemes to complete the $ReKeyGen$ stage remains lengthy. Figure \ref{fig:11}(d) depicts the execution time of the $ReEnc$ stage for each scheme. The MedExChain scheme's execution time is equal to that of the \cite{12} scheme and significantly less than that of the other schemes. This is because both the MedExChain and \cite{12} schemes require only one $P$ operation to complete the $ReEnc$ stage, whereas other schemes require at least four time-consuming operations. Figure \ref{fig:11}(e) illustrates the execution time of each scheme in the $Dec$ stage. The MedExChain scheme's execution time is longer than that of the \cite{21} scheme, equal to that of the \cite{18}, \cite{17}, and \cite{12} schemes, and much shorter than that of the \cite{27} scheme. This is primarily because the MedExChain scheme requires two $P$ operations and one $H$ operation, while the time complexity of the \cite{12} scheme is related to the number of users' attributes, with a significant correlation coefficient, resulting in a much longer execution time for the \cite{12} scheme compared to other schemes. Figure \ref{fig:11}(f) shows the total execution time of each scheme across all stages. The MedExChain scheme's execution time is the same as that of the \cite{12} scheme, smaller than that of the \cite{21} scheme, and much smaller than that of the other schemes. This is mainly because the execution time of each stage in the MedExChain scheme is dominant, making the MedExChain scheme also dominant in total execution time. Overall, the MedExChain scheme offers advantages in computational overhead, making it more competitive than other schemes for IoT devices with limited computational power.\par

\begin{table*}[htbp]
\centering
\renewcommand{\arraystretch}{1.3}
\begin{threeparttable}[c]
\caption{{\MakeUppercase {Comparison of Communicational Overhead}}}
\label{table:communicationalOverhead}
\begin{tabular}{|l|c|c|c|c|c|c|}
\hline
\bfseries Scheme & \bfseries $Key_{DO}$ & \bfseries $Key_{DU}$ & \bfseries $CT$ & \bfseries $RK$ & \bfseries $CT’$ & \bfseries Total (bytes)\\
\hline
\bfseries IBPRE\_CRF \cite{12} & 2$|G_1|$ & 2$|G_1|$ & $|G_1|$+$|G_T|$ & 2$|G_1|$+$|G_T|$ & 2$|G_1|$+2$|G_T|$  & 1,664\\
\hline
\bfseries CDSS \cite{21} & 6$|G_1|$+2$|Z_q|$ & 7$|G_1|$+2$|Z_q|$ & 3$|G_1|$+2$|G_T|$ & 4$|G_1|$+$|G_T|$ & $|G_1|$+3$|G_T|$ & 3,536\\
\hline
\bfseries ABE-IBE \cite{17} & (2$N$+1)$|G_1|$ & 2$|G_1|$ & ($N$+1)$|G_1|$+$|G_T|$ & (4$N$+3)$|G_1|$+$|G_T|$ & 2$|G_1|$+$|G_T|$ & 6,016\\
\hline
\bfseries CP-HAPRE \cite{18} & (2$n$+4)$|G_1|$ & (2$n$+4)$|G_1|$ & (3$N$+2)$|G_1|$+$|G_T|$ & 7$|G_1|$ & 4$|G_1|$+$|G_T|$ & 6,400\\
\hline
\bfseries FABRIC \cite{27} & 6$|G_1|$ & (3$n$+6)$|G_1|$& 9$|G_1|$+$G_T$ & (3$N$+12)$|G_1|$ & (3$N$+9)$|G_1|$+$G_T$ & 10,624\\
\hline
\bfseries \schemeName (Ours) & 2$|G_1|$ & 3$|G_1|$ & 2$|G_1|$+$|G_T|$ & 2$|G_1|$+$|G_T|$ & 2$|G_1|$+2$|G_T|$ & 1,920\\
\hline
\end{tabular}
\begin{tablenotes}
\centering
\item $|G_1|$: Storage overhead of group elements in $G_1$ (128bytes)\quad$|G_T|$: Storage overhead of group elements in $G_T$ (128bytes)
\\$|Z_q|$: Storage overhead of group elements in $Z_q$ (20bytes)\quad
$n$: The number of attributes required by the access policy in ABE\\
$N$: The total number of attributes included in the access policy in ABE
\end{tablenotes}
\end{threeparttable}
\end{table*}

{Additionally, we assessed the variation in execution time for each scheme to complete all user operations as the number of users increased. The experimental results are depicted in Figure \ref{fig:Computational_Overhead1}. As the number of users grows, the time required for each scheme to complete all user operations also increases. Given that the execution time of the MedExChain scheme is shorter at each stage, its computational overhead advantages become more pronounced with an increasing number of users. Figure \ref{fig:Computational_Overhead1}(a) illustrates the execution time of the $KeyGen$ stage for each scheme as the number of users varies. When the number of users is 50, the execution time of the MedExChain scheme is slightly longer than that of the \cite{12} scheme but significantly shorter than that of other schemes, indicating that the MedExChain scheme can efficiently generate key pairs for all users when the number of newly registered users is high. Figure \ref{fig:Computational_Overhead1}(b) shows the execution time of the $Enc$ stage for each scheme as the number of users changes. When the number of users is 50, the execution time of the MedExChain scheme is shorter than that of all other schemes, demonstrating that the MedExChain scheme can swiftly encrypt all medical data when the number of patients is large. Figure \ref{fig:Computational_Overhead1}(c) depicts the execution time of the $ReKeyGen$ stage for each scheme as the number of users varies. When the number of users is 50, the execution time of the MedExChain scheme is equivalent to that of the \cite{18}, \cite{12}, and \cite{21} schemes and considerably less than that of the \cite{17} and \cite{27} schemes. Figure \ref{fig:Computational_Overhead1}(d) illustrates the execution time of the $ReEnc$ stage for each scheme as the number of users changes. When the number of users is 50, the execution time of the MedExChain scheme is equal to that of the \cite{27} scheme and significantly shorter than that of other schemes. The $ReKeyGen$ and $ReEnc$ stages are the most critical stages in the data sharing process. In these stages, the MedExChain scheme exhibits a time advantage when dealing with a large number of users, indicating that it can swiftly complete all data sharing tasks when there are numerous data sharing requests. Figure \ref{fig:Computational_Overhead1}(e) shows the execution time of the $Dec$ stage for each scheme as the number of users varies. When the number of users is 50, the execution time of the MedExChain scheme is longer than that of the \cite{21} scheme but shorter than that of all other schemes. Overall, the MedExChain scheme demonstrates a time advantage, indicating that it can quickly decrypt all medical data and reduce consultation time when the number of patients is large. Figure \ref{fig:Computational_Overhead1}(f) illustrates the total execution time for each scheme as the number of users changes. When the number of users is 50, the execution time of the MedExChain scheme is equivalent to that of the \cite{12} scheme, which is less than that of all other schemes, and its advantages are more evident compared to when there are 10 users. This indicates that the MedExChain scheme can more efficiently complete all tasks in scenarios with a large number of users. Based on the above comparison, the MedExChain scheme is more suitable for medical scenarios with a large number of users than other schemes, enabling IoT devices with limited computing power to handle more user requests.\par}
\begin{figure*}[!t]
    \centering
    \includegraphics[width=17.6cm]{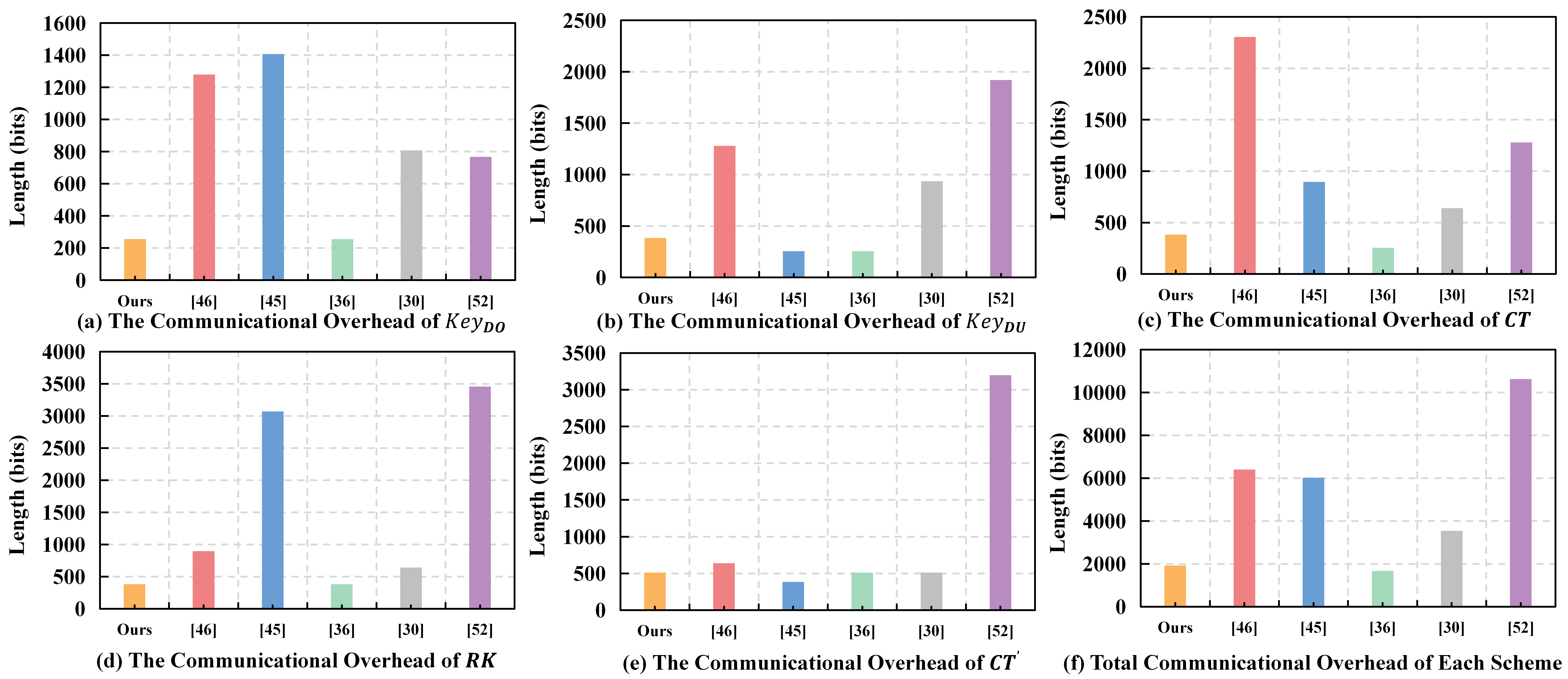}
    \caption{Comparison of Communicational Overhead}
    \label{fig:Computational_Overhead}
\end{figure*}

\subsection{Communicational Overhead}

\subsubsection{Theoretical Analysis} We conducted a theoretical analysis to calculate the communication overhead of each data element in each scheme, including the data owner's key pair ($Key_{DO}$), the data user's key pair ($Key_{DU}$), the ciphertext ($CT$), the re-encryption key ($RK$), and the re-encrypted ciphertext ($CT\prime$). Our analysis focused on the elements within the $G_1$ group, $G_T$ group, and $Z_q$ group contained in each data element.\par
The detailed results of our theoretical analysis are presented in Table \ref{table:communicationalOverhead}. By comparing the number of elements in each data, it is evident that the MedExChain scheme offers advantages in terms of communication overhead. Specifically, in the MedExChain scheme, $Key_{DO}$ consists of elements from two $G_1$ groups, equivalent to \cite{12}, which is significantly smaller than in other schemes. $Key_{DU}$ comprises elements from three $G_1$ groups, which is greater than \cite{17}, \cite{12} but still notably smaller than in other schemes. The $CT$ consists of elements from two $G_1$ groups and one $G_T$ group, which is greater than \cite{17} but slightly smaller than \cite{21} and significantly smaller than in other schemes. The $RK$ consists of two elements from the $G_1$ group and one element from the $G_T$ group, equivalent to \cite{12}, and is significantly smaller than in other schemes. The $CT\prime$ comprises two elements from the $G_1$ group and two elements from the $G_T$ group, which is significantly smaller than \cite{27} and equivalent to other schemes.\par
When the system's security factor is set to 80 bits, the size of elements in the $G_1$ group is 128 bytes, in the $G_T$ group is 128 bytes, and in the $Z_q$ group is 20 bytes. As indicated in the Total (bytes) column of Table \ref{table:communicationalOverhead}, we calculated the theoretical total communication overhead for all data elements in each scheme. The MedExChain scheme's total overhead is slightly larger than \cite{17} but significantly smaller than in other schemes. Based on this analysis, it is evident that the MedExChain scheme offers advantages in communication overhead, making it a more suitable choice for IoMT devices with limited storage space and network bandwidth.\par 
\subsubsection{Experimental Measurement} In the experimental setup described, we implemented the code for each scheme and measured the size of each data element by determining its byte length. The experimental results, depicted in Figure \ref{fig:Computational_Overhead}, align with the theoretical analysis, demonstrating that the MedExChain scheme exhibits a favorable position in terms of communication overhead. Figure \ref{fig:Computational_Overhead}(a) illustrates the length of $Key_{DO}$ for each scheme. The length of $Key_{DO}$ in the MedExChain scheme is equivalent to that in the \cite{12} scheme and shorter than in other schemes, as $Key_{DO}$ in both the MedExChain and \cite{12} schemes consists of elements from two $G_1$ groups, whereas $Key_{DO}$ in other schemes includes at least six $G_1$ groups. Figure \ref{fig:Computational_Overhead}(b) presents the length of $Key_{DU}$ for each scheme. The length of $Key_{DU}$ in the MedExChain scheme is marginally greater than that in the \cite{17} and \cite{12} schemes but shorter than in other schemes, due to the MedExChain scheme's $Key_{DU}$ containing one additional element in the $G_1$ group compared to the \cite{17}, \cite{12} schemes, while other schemes' $Key_{DU}$ includes at least twice as many elements. Figure \ref{fig:Computational_Overhead}(c) shows the length of $CT$ for each scheme. The length of $CT$ in the MedExChain scheme is slightly greater than that in the \cite{12} scheme but shorter than in other schemes, as the $CT$ in the MedExChain scheme includes one additional element in the $G_1$ group compared to the \cite{12} scheme, and other schemes' $CT$s contain at least twice as many elements. Figure \ref{fig:Computational_Overhead}(d) depicts the length of $RK$ for each scheme. The length of $RK$ in the MedExChain scheme is identical to that in the \cite{12} scheme and shorter than in other schemes, as $RK$ in both the MedExChain and \cite{12} schemes consists of elements from two $G_1$ groups and one $G_T$ group, whereas $RK$ in other schemes includes at least four $G_1$ groups and one $G_T$ group. Figure \ref{fig:Computational_Overhead}(e) illustrates the length of $CT\prime$ for each scheme. The length of $CT\prime$ in the MedExChain scheme is slightly greater than that in the \cite{17} scheme, equal to that in the \cite{12} and \cite{21} schemes, and significantly shorter than that in the \cite{18} and \cite{27} schemes. This is because the MedExChain scheme's $CT\prime$ contains one less element compared to the \cite{18} scheme, and the number of elements in the $G_1$ group in the $CT\prime$ of the \cite{27} scheme is correlated with the number of attributes in the access policy, resulting in a larger $CT\prime$ length for the \cite{27} scheme. Figure \ref{fig:Computational_Overhead}(f) shows the total length of all data elements for each scheme. The total length of data in the MedExChain scheme is slightly greater than that in the \cite{12} scheme but shorter than in all other schemes, as the length of certain data elements in the MedExChain scheme is slightly greater than in the \cite{12} scheme, while the data length in the MedExChain scheme is generally shorter compared to other schemes. Overall, the MedExChain scheme demonstrates advantages in communication overhead, effectively reducing the time and traffic consumption of IoMT devices during data transmission.

\subsection{Performance of Blockchain System}
In this section, we deployed each scheme, simulated cross-chain data sharing between two blockchains, and measured the actual performance of systems implementing these schemes in handling cross-chain requests. We then conducted analysis and comparisons accordingly.
\begin{figure}[!t]
    \centering
    \includegraphics[width=3.50in]{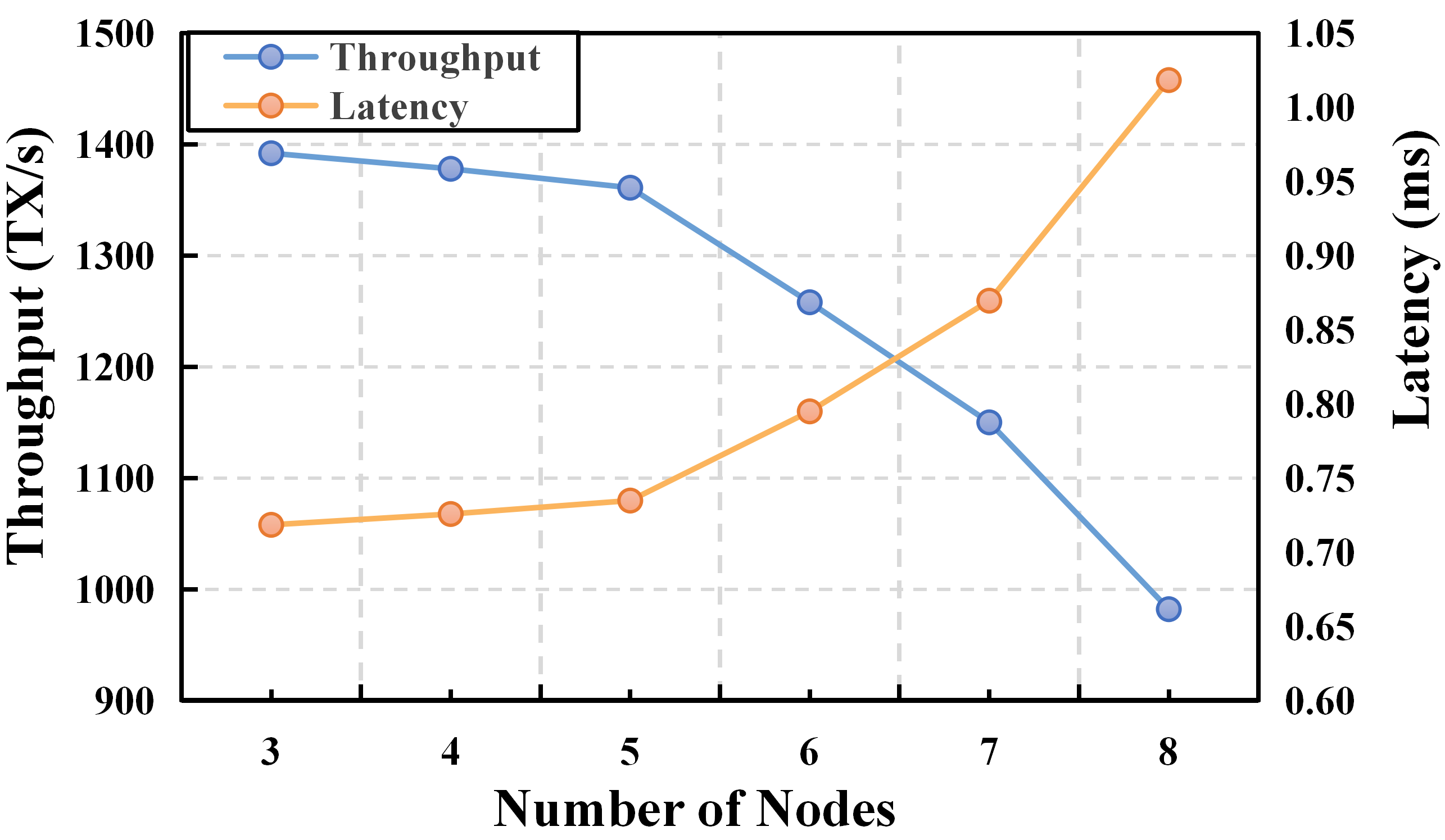}
    \caption{System Performance of Multi-Nodes}
    \label{fig:nodes}
\end{figure}

{\subsubsection{Relationship Between System Performance and Number of Blockchain Nodes}
Considering the impact of the number of nodes on the performance and security of blockchain systems, we first investigated the relationship between system performance and the number of blockchain nodes. Due to performance constraints, we configured blockchains with 3, 4, 5, 6, 7, and 8 nodes, respectively, and deployed a simple smart contract to store a piece of information on each blockchain. By concurrently sending 100,000 requests to each blockchain, we measured the throughput and latency of each system.
As depicted in Figure \ref{fig:nodes}, with an increase in the number of blockchain nodes ($n$), the system's throughput decreases while latency increases. This is attributed to the increased time and communication overhead required for nodes to reach consensus, which diminishes throughput and elevates latency.
In blockchain systems, a higher number of nodes equates to more distributed copies, enhancing system security. Given the diminishing rate of system throughput observed in Figure \ref{fig:nodes}, when the number of nodes exceeds 5, a balance between system performance and security is achieved. Consequently, we selected 5 nodes for subsequent experiments.}

\subsubsection{System Performance of Schemes}
We constructed two blockchains in the experimental environment and deployed each scheme on these blockchains. Following deployment, we performed system initialization, key pair generation, and ciphertext generation operations. Subsequently, we instructed Blockchain B to send a request to obtain the re-encrypted ciphertext and measured the system's throughput and latency by calculating the time from request transmission to receipt of the re-encrypted ciphertext. During actual measurements, Blockchain B concurrently sent 10,000 requests to Blockchain A.
The results are illustrated in Figure \ref{fig:systemPerformance}. By comparing throughput and latency, it is evident that the MedExChain scheme is on par with \cite{12}, superior to \cite{18} and \cite{21}, and significantly outperforms \cite{17} and \cite{27}. This demonstrates that the MedExChain scheme offers overall performance advantages. This is primarily due to the simplicity and efficiency of operations in the $ReKeyGen$ and $ReEnc$ stages, minimal communication overhead per data unit, and short data transmission times. These characteristics enable the MedExChain scheme to operate effectively and efficiently, even with limited computing and storage resources on IoMT devices.}

\begin{figure}[!t]
    \centering
    \includegraphics[width=3.50in]{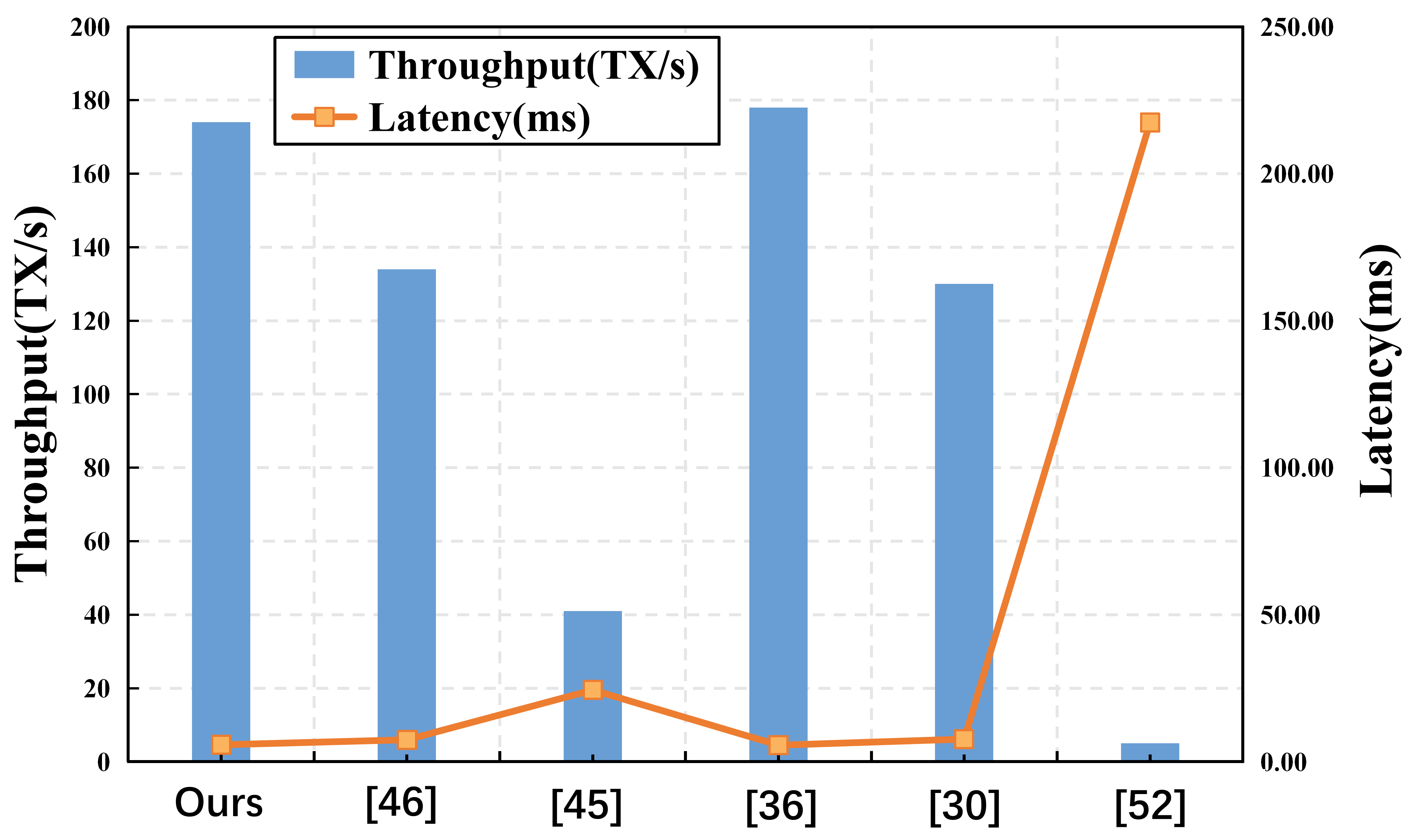}
    \caption{System Performance in Different Schemes}
    \label{fig:systemPerformance}
\end{figure}

\section{Conclusion}\label{CON}
This paper delves into the cross-chain sharing of PHR within intelligent medical systems. The proliferation of AI technology has underscored the significance of cross-system PHR utilization, yet disparate system password mechanisms and security concerns impede effective sharing across various medical systems. To address these challenges, we introduce a novel cross-heterogeneous blockchain scheme (MedExChain), which facilitates secure PHR sharing among blockchains employing diverse cryptographic systems. This scheme enables PHR sharing via smart contracts, even in scenarios where IoMT devices exhibit limited performance, thereby mitigating computational and storage demands. MedExChain safeguards the data sharing process against both internal and external threats while demonstrating robust performance. Future research will concentrate on refining the consensus mechanism between heterogeneous blockchains during cross-chain interactions, with the aim of enhancing the security and efficiency of data sharing.

\section*{Acknowledgment}
This work was supported by the National Key Research and Development Program of China (2021YFF1201102).

\renewcommand{\bibfont}{\footnotesize} 
\printbibliography

\vspace{-1.5cm}
\begin{IEEEbiography}[{\includegraphics[width=1in,height=1.25in,clip,keepaspectratio]{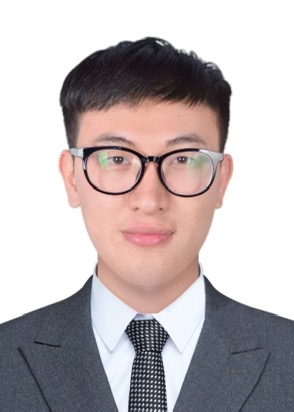}}]{Yongyang Lv}
received a Master's degree from Tiangong University in 2021 and is currently pursuing a doctoral degree at Tianjin University. His research focuses on network security, data security, and blockchain security.
\end{IEEEbiography}

\vspace{-1.5cm}
\begin{IEEEbiography}[{\includegraphics[width=1in,height=1.25in,clip,keepaspectratio]{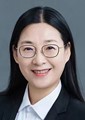}}]{Xiaohong Li}
(Member, IEEE) received the Ph.D. degree in computer application technology from Tianjin University in 2005. She is currently a Full Tenured Professor with the Department of Cyber Security, College of Intelligence and Computing, Tianjin University. Her research interests include knowledge engineering, trusted computing, and security software engineering.
\end{IEEEbiography}

\vspace{-1.5cm}
\begin{IEEEbiography}[{\includegraphics[width=1in,height=1.25in,clip,keepaspectratio]{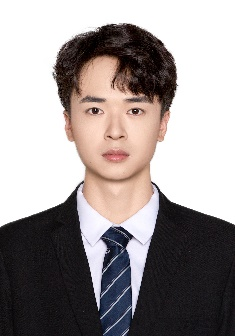}}]{Kui Chen}
is currently pursuing a bachelor's degree at Tianjin University and has been accepted into the Master's program at the School of Software Technology, Zhejiang University. His research interests encompass blockchain technology and software engineering.
\end{IEEEbiography}
\vspace{-1.5cm}
\begin{IEEEbiography}[{\includegraphics[width=1in,height=1.25in,clip,keepaspectratio]{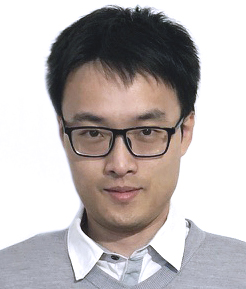}}]{Zhé Hóu}
is a Senior Lecturer at Griffith University, Australia. He obtained his Ph.D. degree from the Australian National University in 2015. His research mainly focuses on automated reasoning, formal methods, AI, quantum computing and blockchain.
\end{IEEEbiography}
\vspace{-1.5cm}
\begin{IEEEbiography}[{\includegraphics[width=1in,height=1.25in,clip,keepaspectratio]{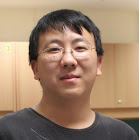}}]{Guangdong Bai}
(Member, IEEE) received the B.S. and M.S. degrees in computing science from Peking University in 2008 and 2011, respectively, and the Ph.D. degree in computing science from the National University of Singapore in 2015. He is currently a Associate Professor with The University of Queensland. His research interests include cyber security, software engineering, and machine learning.
\end{IEEEbiography}
\vspace{-1.5cm}
\begin{IEEEbiography}[{\includegraphics[width=1in,height=1.25in,clip,keepaspectratio]{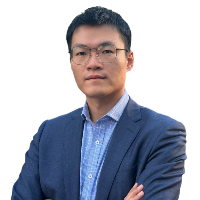}}]{Ruitao Feng}
is a Lecturer at Southern Cross University, Australia. He received the Ph.D. degree from the Nanyang Technological University. His research centers on security and quality assurance in software-enabled systems, particularly AI4Sec \& SE. This encompasses learning-based intrusion/anomaly detection, malicious behavior recognition for malware, and code vulnerability detection.
\end{IEEEbiography}

\end{document}